\documentclass[5p,times]{elsarticle}

\usepackage{lineno,hyperref}
\modulolinenumbers[5]

\journal{ArXiv}

\usepackage{amsmath}
\usepackage{url}
\usepackage{booktabs}
\usepackage{geometry}
\usepackage{CJK}
\usepackage{multirow}

\usepackage{graphicx}
\usepackage{epstopdf}
\usepackage{algorithm}
\usepackage{algorithmic}

\usepackage{amsmath}
\usepackage{subfigure}
\usepackage{multirow}
\usepackage{graphicx}
\usepackage{times}
\usepackage{subfigure}         
\usepackage{natbib}
\usepackage{amssymb,amsmath}
\begin{document}

\begin{frontmatter}

\title{An Optimized BBR for Multipath Real Time Video Streaming}
\author{Songyang Zhang\fnref{myfootnote}}
\address{School of Computer Science and Engineering, Northeastern University, China}
\fntext[myfootnote]{sonyang.chang@foxmail.com}
\author{Weimin Lei\fnref{myfootnote1}}
\address{School of Computer Science and Engineering, Northeastern University, China}
\fntext[myfootnote1]{leiweimin@ise.neu.edu.cn}
%
%
%

\begin{abstract}
The multipath transmission scheme can work as an effective way to provide better quality of experiments to end users. Two key research points in the multipath real time video transmission context are congestion control and packet scheduling. As Utility maximization theory shows, to provide better satisfaction to end users is to provide higher throughput and lower transmission delay. The congestion control is responsible to converge to the maximum available bandwidth and avoid leading the network into congestion. A delay response BBR (Delay-BBR) algorithm optimized for real time video transmission is proposed, and the main idea is to reduce sending rate when the link delay has exceeded certain threshold to let the intermediate routers drain the occupied buffer. It can achieve better transmission delay and lower packet loss rate compared with QUIC-BBR and WebRTC-BBR by experiment. And a packet scheduling algorithm induced from Utility maximization theory works on top of the congestion control algorithm is tested and achieves lower frame delivery delay further compared with benchmark algorithms.
\end{abstract}


\end{frontmatter}

\section{Introduction}
According to a recent report \cite{cisco-report}, video currently accounts more than 70\% of all internet traffic. The Quality of Experience (QoE) of video streaming is still a main concern to video based applications developers and service provider. Video based applications may suffer from packet loss, restricted bandwidth and re-buffering event for the Internet best effort service mode, even though the internet infrastructure has achieved huge advance. The rate of video streaming in internet is highly dynamic due to bandwidth resource competition. The congestion control algorithm in Transmission Control Protocol (TCP) would send more packets to probe bandwidth during the congestion avoidance phase would lead the network into congestion status unavoidable. 

Currently, there are two widely applied technologies DASH (dynamic adaptive streaming over HTTP) and WebRTC \footnote{https://webrtc.org/} to optimize the QoE for video transmission at the application layer. The DASH is mainly applied by video content providers e.g., YouTube and NetFlix, while WebRTC is used for real time video transmission to build interactive multimedia application. The application scenarios are different, but their final goals are same, to provide better QoE based on current network status. In DASH systems, a duration of video content is encoded into different bitrate chunks, a client can choose the appropriate chunk based on the estimated available bandwidth to avoid playback stalls and rebuffing events. WebRTC controls the bitrate and resolution of encoder by implementing congestion control algorithm on top of UDP. Both technologies will reduce video bitrate when the network link falls into congestion status. The reduced video bitrate would produce perceptible video quality loss to end user. The increased end to end latency due to congestion leads the receiver buffer underrun and the video rendering process stalls. Such factors are quite harmful to QoE. 

And there exists another solution to provide better QoE by exploiting multipath transmission scheme, which has been a hot topic in academia area. Today’s Networks are becoming multipath. Mobile devices equipped with multi-homed interfaces make the concurrent access to heterogeneous networks (WIFI and 4G) possible. The datacenter network provides redundant paths for high availability concern. Enterprises would choose a secondary Internet service provider (ISP) to reduce internet outrage risk. And the multipath transmission scheme brings out many advantages such as providing more bandwidth resource, accelerating flow completion time and improving robustness in case of single path route failure. There are two transport layer protocols MPTCP and CMT-SCTP designed to take advantage of multipath transmission scheme. The two protocols are mainly applied for bulk data transmission and provide reliable packets transmission mechanism. A reliable transmission protocol would delay the delivery all the subsequent packets until the lost packet is retransmitted, which will introduce considerable latency. For applications that have stringent delay requirement, especially for interactive video streaming, UDP is a more preferable choice. For real time video transmission application, skipping a none key frame may not bring out obvious quality deterioration as latency do. There are not too many research works on multipath transmission scheme for real time video traffic that works on top of application layer.

Two key research points on multipath transmission scheme are congestion control and packet scheduling. The goal of congestion control in MPTCP context is to provide protocol friendliness when the multipath session shares bottleneck with single path session \cite{rfc6356}. And there are several fruitful results on this area such as LIA \cite{rfc6356}, OLIA \cite{Khalili2013MPTCP}, wVegas \cite{Cao2012Delay}. There are a lot of works related to packet scheduling algorithm in MPTCP \cite{Frommgen2016ReMP, Guo2017Accelerating, Lim2017ECF, Shi2018STMS, Shreedhar2018QAware} context. In the MPTCP context, the goal of the scheduling algorithms is to improve the throughput and minimize flow completion time by taking the concept of delivering packets out of order to achieve in order arriving.  Due to the heterogeneity of the routing paths, the packets scheduled to different paths may arrive out of order at receiver, which increases the buffer occupancy and may cause the head of line blocking \cite{Ferlin2016BLEST} phenomena. Once the buffer occupancy exceeds the pre-allocated memory at the transportation layer, the following incoming packets will be dropped. The sender will half its congestion window in loss base congestion control algorithm even the packet loss is not caused by network congestion, which leads the throughput of MPTCP is far from being optimal.

But the multipath congestion algorithms are an improvement to Additive Increase Multiplicative Decrease (AIMD) law proposed by Jocobson \cite{jacobson1988congestion} in the multipath context. The congestion window is halved when link congestion event is detected, which makes the packet sending rate show saw-tooth feature. Such mechanism causes the video encoder instability and provides unsufferable experience to end user. The congestion control algorithm in WebRTC prefers a rate back off factor $0.8$ when link is in congestion in order to maintain stable video frame generating bitrate. And the purpose of the packets scheduling algorithm for real time traffic is to provide better QoE by taking advantage of path heterogeneity. The packet scheduling for real time video traffic is not suffering from the suboptimal throughput due to the buffer overflow at receiver side as in MPTCP. The reason is that buffer can be allocated dynamically at the application layer, thus buffer overflow will not happen. 

In this work, a multipath transmission mechanism for real time video traffic taking both congestion control and packet scheduling into consideration is proposed. Inspired by the excellence performance of the congestion control algorithm BBR \cite{cardwell2016bbr}, a delay response BBR (Delay-BBR) optimized for real time video transmission is proposed to overcome its bandwidth aggressiveness which causes high packet loss rate and transmission latency in its implementation in QUIC and WebRTC. By actively reducing sending rate to let the intermediate routers drain the queued buffer when the delay signal exceeds defined threshold, Delay-BBR can achieve obvious lower packet loss rate and lower transmission latency. A packet scheduling algorithm working on the proposed congestion control algorithm is proposed, which is based on utility maximization theory and takes the local queued packets length, available bandwidth and transmission delay into consideration. Simulation results show the congestion control algorithm can make a fast convergence the available bandwidth, maintain rate stability and guarantee well fairness when competing resource with other flows exploiting the same congestion control algorithm. The packet scheduling algorithm achieves the lowest average frame delay when compared with the benchmark algorithms.  

The rest of this paper is organized as follows. Section 2 provides a brief review of related works on congestion control algorithms especially for real time video traffic and packet scheduling algorithms. The propose multipath transmission framework for real time video and the theory to provide better satisfaction are shown in Section 3. The implementation detail on congestion control algorithm and packet scheduling algorithm are described in Section 4. The performances evaluation and comparison are presented on Section 5. Section 6 is the conclusion and some directions on future research.
\section{Related work}
In single transmission scheme, the main component on transport layer is congestion control. As pointed by \cite{ietf-rmcat-require}, all the flows across internet should implement congestion control scheme for internet congestion avoidance and promote fair bandwidth occupation. Without the restrictions of the congestion control algorithm, the network user would send packets in selfish manner and network would have the risk to fall into collapse. The seminal work \cite{jacobson1988congestion} in TCP congestion avoidance area, Jocobson proposed to regulate TCP sending rate according to the law of additive increase and multiplicative decrease (AIMD), which makes the network congestion control an unfading topic in computer networks research. Most of the later research works such as Bic \cite{xu2004binary}, Cubic \cite{ha2008cubic}, were proposed to improve TCP performance and adapted the basic AIMD control law to different network environment. The main change is bandwidth probe mechanism during the congestion avoidance phase to make the control algorithm converge quickly to the fair sharing bandwidth.

The main purpose of current proposed multipath congestion control algorithms such as LIA \cite{rfc6356}, OLIA \cite{Khalili2013MPTCP}, and wVegas \cite{Cao2012Delay} is to promote the deployment of MPTCP by providing high throughput and guarantee protocol friendliness when the multipath TCP session competing resource with single TCP session. These algorithms couple the sub-flows into aggregation to achieve friendliness and do not change the basic working mechanism of AIMD.

The sending rate of TCP will be sharply reduced in face of link congestion. If such mechanism is directly applied for real time video traffic, it will cause the instability of video encoder. And loss based congestion control algorithm tend to occupy much buffer resource in the intermediate routers during the bandwidth probe phase and results long transmission delay, which is notorious for bufferbloat \cite{Staff2012Bufferbloat}. These factors make it not appropriate to be applied for real time video transmission.

In consideration of the above mentioned drawbacks of TCP congestion control algorithm, there have been a trend to deploy a congestion control algorithm at the application layer, specifically optimized for real time video transmission in recent years. The IETF has initiated The RTP Media Congestion Avoidance Techniques (RMCAT) Working Group to develop congestion standards for interactive real-time media. And there are three congestion control drafts under this working group, namely, GCC \cite{carlucci2016analysis}, NADA \cite{zhu2013nada}, SCReAM \cite{rfc8298}. Three algorithms take delay signal to indicate whether the link is in congestion status in order to improve the packet transmission latency. GCC takes one way delay gradient for congestion signal, NADA takes an aggregated delay signal and SCReAM takes one way delay for rate control. The rate control of these algorithms is to converge to the available bandwidth as close as possible while keeping a low end to end delay.

In multipath transmission scheme, a packet scheduling algorithm functions to provide better performance. And the packet scheduling algorithm is depended on the congestion control algorithm, which provides the basic information on the link available bandwidth. The main purpose of packet scheduling algorithm \cite{Kuhn2014DAPS} in MPTCP context is to avoid buffer blocking at the receiver. The recent research works on packets scheduling algorithm in multipath transmission context are briefed in the following. 

DEMS \cite{Guo2017Accelerating} is a chunk based scheduling algorithm for MPTCP to achieve simultaneous sub-flow completing time. And the chunk is defined by application layer, and strategically split according to bandwidth and one way delay to be sent on different path in opposite direction. STMS \cite{Shi2018STMS} schedules packets with smaller sequence number to fast path while sending packets with larger sequence number on slow path, through which the out order arrival is reduced. In \cite{Shreedhar2018QAware}, the authors designed a cross layer packet schedule algorithm which takes the packet queue length and packet pending delay in NIC into consideration. On a packet comes, the scheduler would choose the NIC interface that can deliver the packets earliest. 

OMS \cite{Cetinkaya2004Opportunistic} proposes to schedule the incoming packet to the path with the minimal arriving delay. The packet arriving delay is estimated by link capacity and buffered queue length. In real network situation, the link capacity can be measured by packet pair like technology \cite{Ribeiro2003pathchirp}, but the link queue occupation size and the number of flows through the same routing path are quite hard to estimate. In his simulation results, the queue occupation following asymptotically Weibull distribution \cite{Norros1994storage} is assumed and the path has translational Brownian motion (tBm) traffics with prior known mean rate, variance coefficient and Hurst parameter.

In \cite{Zhu2007Rate}, a rate allocation scheme in heterogenous access networks is proposed, in which the rate control is combined with packet scheduling. A rate distortion is introduced to work as the optimization function. The rate distortion equation is related to video encode rate and packet loss rate. The packet loss rate is modeled from M/M/1 queue. Increase the allocated rate on path $i$ would increase the packet arriving delay. When a packet arriving delay is larger than the maximum frame playout deadline, the packet is count as loss packet. Whether the M/M/1 queue theory can reflect the true status of network link is quite questionable.

In \cite{Bui2010Markovian}, a Markov Decision Process is applied for packet distribution over multipath overlay network. The incoming packets are segmented into fixed size bins before delivering to traffic distributor to choose the path. The objective of the MDP approach is to maximize the average reward over a short time period, or to minimize the packet transfer time. The immediate reward of a specific path is defined as the average packet inter-departure time and propagation delay.

EDCLD \cite{Prabhavat2011Effective} takes a hybrid model to compute the path delay, which is denoted as cost function. The cost function is modeled from Poisson traffic and unknown traffic. Its goal is to minimize the path cost variation though packet scheduling. Increase the packets splitting ratio on path with smaller cost and decrease it on path with larger cost and finally the balance costs are reached on all paths.

SFL \cite{Wu2015low} splits the large video frame into sub-frames level to optimize video delivery delay.  The end to end delay is estimated by M/M/1 queue model. The watering filling algorithm is applied in SFL, which schedules packets to specific path according to available bandwidth and delay to make the all the paths have equal water bucket depth.

Through a brief overview of the proposed algorithms on multipath packet scheduling, nearly all of them share a common idea that the optimization is achieved by scheduling packet to path with less cost and the core differences of these works are the definition of routing path cost which may be expressed by delay, packet loss rate, or both. And we have to admit here, this main idea of proposed packet scheduling algorithm also belongs to such kind in essence. But we argue here, due to the different definition of cost function, the performance of these algorithms shows diversity in simulation.
\section{Framework}
\subsection{System overview}
\begin{figure*}
\centering
\includegraphics[width=7in]{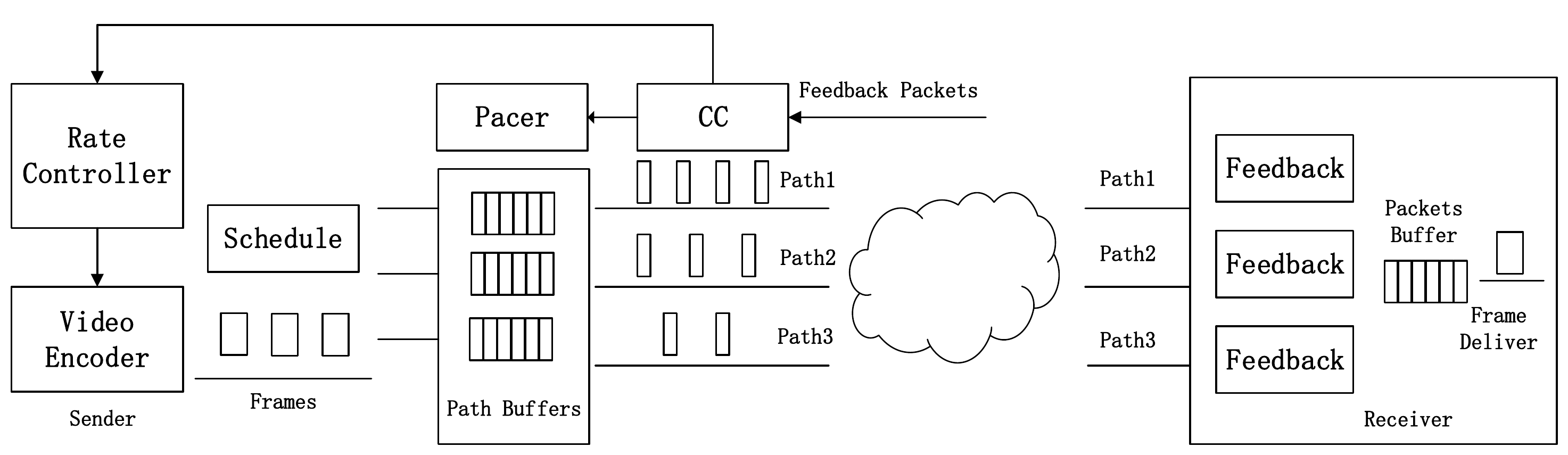}
\caption{The designed multipath transmission framework}
\label{Fig:system}
\end{figure*}
The multipath transmission framework applied for real time video traffic in this work is shown in Figure \ref{Fig:system}. The real time video traffic is usually transported via RTP (Real Time Protocol) over UDP for latency consideration. Considering the large scale and popular of real time video traffic, a congestion controller (CC) module must be introduced to guarantee a fair share of routing path capacity when competing bandwidth with other flows and avoid leading the network into congestion. On each sub-path, a congestion control module is implanted, and only one is showed in the diagram due to the limited of space. 

And the pacer module functions to inject local buffered packets into the network with the rate determined by the congestion control module. Since the video frame is generated at some fixed interval, the traditional congestion window based packet sending mode would send the packet in burst, which will cause the packets queued in the intermediate routers and extra delay, and is not quite appropriate for video transmission. There is a tendency to implement the pacing function in current newly proposed congestion algorithms e.g. BBR and GCC to send the packets out evenly, and to match the available bandwidth. There are some references \cite{Aggarwal2000Understanding, Wei2006TCP} argue about the beneficial of pacing. 

The schedule module is responsible for splitting the incoming frame into packets and schedules these packets to different path buffer according to specific cost function. If the packets incoming rate of a path exceeds the pacing rate, the length of local buffer would increase and the latency of buffered packet delivered to the receiver would increase too. 

The rate control module adjusts bitrate of the video encoder according to the available bandwidth of all its sub-paths to make sure the video data generating rate to match the total available throughput. 

At the receiver side, each sub-flow would feedback the packets received information for the sender. Such information is fed to the congestion control to determine the packet sending rate. The packets sent on different path will arrive out of order at the receiver side, hence a buffer is allocated at the receiver. When a completed frame is available, the frame will be delivered to upper layer for video rendering.

The multipath transmission protocol applied in simulation will be briefed. The protocol is taken reference from razor project \footnote{https://github.com/yuanrongxi/razor}. The incoming video frame is packed by segment. Apart from the video data, some extra information is included in the segment header: The frame id (fid), the total packed segments of a frame, the segment index in a frame. The segment index is used for putting the received segment in the right position in the receiver buffer and to decide whether the data of the segment has already received. There is a tendency to separate packet number from transport sequence number in newly proposed protocol such as QUIC. RTP has an extension proposal \cite{holmer2015rtp} for such mechanism which has already implemented in WebRTC.

Even the RTP/RTCP protocol format is not implemented in simulation, it is easy to substitute the implemented protocol in RTP format.
\subsection {Problem Formulation}
A utilization function $U$ is introduced to describe the user’s satisfaction of a service, which is usually assumed as strictly concave and increasing function. This optimization problem is to maximum the utilization under the subject of link available bandwidth. The network system is modeled as a set of $L$ routing paths. $S$ denotes the number of network users. A user $s$ can exploit multiple routing path to build concurrent connections. $x_{s,l}$ is the packet sending rate of user $s$ in path $l$, which of course is zero if the path is not used by $s$. The total rate of user $s$ is denoted as $y_s$ and $y_s=\sum_{l\in L}{x_{s,l}}$. The bottleneck capacity (the narrowest link) of path $l$ is $c_l$. User $s$ will get a satisfaction or welfare measured by $U(y_s)$ when transmitting packet with rate $y_s$. The goal is to maximum of aggregate Utility function under the constraint of path capacity. Thus, the following optimization problem is obtained:
\begin{equation}
\begin{aligned}
\max \quad & \sum_{s\in S}U(y_s)\\
\textrm{s.t.} \quad &\sum_{s\in S}x_{s,l}\leq c_l\\
\end{aligned}
\end{equation}

In computer network, a global resource allocation solution is quite hard considering the large scale of current network systems and such problem is always solved by distribution method. According to the Lagrange dual technology introduced by Kelly \cite{Kelly1998Rate},  the Lagrangian is defined:
\begin{equation}
\begin{aligned}
L(y_s,\lambda)&=\sum_{s\in S}U(y_s)+\sum_{l}\lambda_{l}({c_{l}-\sum_{s\in S}x_{s,l})}\\
&=\sum_{s\in S}U(y_s)-\sum_{l}\lambda_{l}(\sum_{s\in S}x_{s,l})+\sum_{l}\lambda_{l}c_{l}\\
&=\sum_{s\in S}U(y_s)-\sum_{s\in S}\sum_{l}x_{s,l}\lambda_{l}+\sum_{l}\lambda_{l}c_{l}
\end{aligned}
\end{equation}

The parameter $\lambda_{l}$ is the shadow price introduced by Kelly \cite{Kelly1998Rate} of path $l$. More specifically, the shadow price is congestion signal  feedback by network system, which may be increase delay or packet loss event due to channel overuse. According to previous research works \cite{Kelly1998Rate}\cite{Low1999Optimization}\cite{Cao2012Delay}, the dual function of the original problem is:
\begin{equation}
\label{eq:dual}
\min_{\lambda}D(\lambda)
\end{equation}
The function $D(\lambda)$ is defined as:
\begin{equation}
D(\lambda)=\max L(y_s,\lambda)=\sum_{s\in S}L_s(\lambda)+\sum_{l}\lambda_{l}c_{l}
\end{equation}
Where
\begin{equation}
\label{eq:subproblem}
L_s(\lambda)=\max{U(y_s)-\sum_{l}x_{s,l}\lambda_{l}}
\end{equation}

Equation \eqref{eq:subproblem} is defined as aggregate surplus as in \cite{Paganini2009Unified}. To pursuit the maximization of aggregate surplus is to increase the packet sending as highly as possible or choose the routing path with the minimal routing cost. Due to limit of the link capacity, the user sends packet with fast rate will lead the congestion of the network and increase the cost in the second part of the equation and lead aggregate surplus decrease. Hence, the solution to maximize aggregate surplus is to minimize the routing cost under the constrict of the congestion control algorithm. Inspired from BBR which achieves high bandwidth utilization, an improve version optimized for real time video transmission is implemented in this work to overcome some drawbacks of original BBR.  The congestion control achieves a fair share of the available bandwidth. So, to minimize the routing costs is the task of the packet schedule algorithm.  And then the following analysis is mainly focused on packet scheduling algorithm.

To make the utilization theory can be well applied in the multipath packet schedule problem, a minor modification is made related to equation \eqref{eq:subproblem}.
\begin{equation}
\begin{aligned}
\label{eq:schedule}
\max \quad &U(x)-\sum_{p}\lambda_{p}x_p\\
\end{aligned}
\end{equation}
where, $p$ is the path index, $\lambda_{p}$ is path price, and $x_p$ is the packets schedule rate. The packets generating rate is x, $x_p=\alpha_{p}x$. We further assume that even in a single path transmission scheme, the exploited path can guarantee satisfaction to some extent and is usable for video transmission. That means there exists a rate $x_p>0$ to make $U(x_p)-\lambda_{p}x_p>0$. And multipath transmission can provide better satisfaction than a single path transmission. If not, the multipath transmission will show no advantages, which is counter-intuitive.  But an ill designed scheduling algorithm will make a sub-optimal solution to \eqref{eq:schedule}. It will increase the local packet buffer and packet delivery latency, which can be considered as ``congestion''.

The aggregate cost of multipath sesssion is denoted by $q$, which can be interpreted as the expected cost of a single packet scheduled through multipath \eqref{eq:single-cost}.
\begin{equation}
xq=\sum_p{x_{p}\lambda_{p}}
\end{equation}
\begin{equation}
\label{eq:single-cost}
q=\sum_{p}{\lambda_{p}\alpha_{p}}
\end{equation}

The link cost $\lambda_{p}$ can be expressed by \eqref{eq:path-cost}. $d_p$ is the propagation delay. $d_q(t)$ is the queue delay introduced by the packets buffered at the intermediate router. And the third term is the packet queue delay at local buffer, $\boldsymbol 1()$ is the step function. The dynamic of the link cost $\lambda_p$ can be denoted by \eqref{eq:cost-dynmic}, if the dynamic of delay introduced at intermediate router is ignored. Once the scheduling rate $x_p(t) $ exceeds the path sending rate $c_p(t)$, the extra send packets would be buffered at the sender sider which would increase the end to end delay $\dot \lambda_p(t)>0$.
\begin{equation}
\label{eq:path-cost}
\lambda_p(t)=d_p(t)+d_q(t)+\int{\frac{(x_p(t)-c_p(t))\boldsymbol 1(x_p(t)-c_p(t))}{c_p(t)}}\mathrm{d}t
\end{equation}
\begin{equation}
\label{eq:cost-dynmic}
\dot \lambda_p(t)=\frac{x_p(t)-c_p(t)}{c_p(t)}
\end{equation}

As for the first term of the equation \eqref{eq:schedule}, there is quite little impact a packet schedule algorithm can wield. The main focus of packet schedule algorithm is to minimize the second term. We further require that a packet with size equal or less than the Maximum Transfer Unit (MTU) is the meta packet that is none separable. The aggregate cost equation \eqref{eq:single-cost} can be rewrite as the following:
\begin{equation}
\label{eq:re-cost}
q_{t}=\boldsymbol{\lambda_{p,t}}\boldsymbol{\alpha_{p,t}}
\end{equation}

Here, the vector $\boldsymbol{\alpha_{p,t}}$ works as an indicator to show which path of an incoming packet in time $t$ should be scheduled. It's obvious that scheduling a meta packet with the goal of minimizing cost is to choose the path with the minimal cost in current time. The proposed solution is to increase the packets scheduling rate to the path with smaller cost and decrease it on the path with larger cost. Assuming the frame generating rate is set as the sum of available bandwidth of all the sub-paths $x=\sum_{p}{c_p}$, it means a single path cannot sending all the packets within the deadline requirement. There exists a time point that makes a packet arriving latency over 500 milliseconds, which causes un-sufferable QoE. Scheduling more packets to the minimal cost path will further increase the path cost as equation \eqref{eq:cost-dynmic} indicates. Such method can strike a balance in terms of link cost and adapts well when path available bandwidth changes. In Lao Tzu's words, the way of heaven, reduces the surplus to make up for scarcity.
\section{Implementation}
First, the implementation of the congestion control algorithm will be described in this part. Then, to verify the performance of the proposed packet scheduling method, several benchmark algorithms are implemented based on our experiment framework. The multipath real time video transmission frame work shown in Figure \ref{Fig:system} are implemented on the ns3 simulation platform. 
\subsection{The implementation detail on congestion control algorithm}
The packet scheduling algorithm should work on top of congestion control algorithm. Without the congestion control module, the endpoint would not know the appropriate number of packets to be sent into network link. In an early stage during the preparation of this work, GCC algorithm is applied and we observed in simulation that the rate in initial phase ramps up slowly in underload path and the total throughput is sub-optimal. The reason is analyzed here. The video frames are captured from camera at setting intervals. To deliver the packets to the earliest path according to the packet schedule algorithm would make the other path underload. And the underload path may have not enough packets to probe the available bandwidth When combined with congestion control algorithm, which results quite low throughput on such path.
 
The most recent proposed congestion control algorithm is BBR, which has remarkable performance in bandwidth utilization. Basically, BBR is claimed as a congestion based congestion control algorithm, with the goal to reach the optimal control point, namely, to achieve the maximum throughput while minimizing end to end latency at the same time. Its sending rate is adjusted according to the probed available bandwidth, and it is not response to link delay increase or packet loss event. And BBR flow show aggressiveness by insisting the maximum probed bandwidth when competing bandwidth with other BBR flows, and its optimal control point is thus deviated. Even though it can approach quite high throughput, the packet loss rate and transmission latency are quite high. 

There are four states StartUp, Drain, ProbeBW, and ProbeRTT during the data transmission in BBR. The Startup and Drain states are used at a session start. At the Startup phase, the endpoint uses a high gain $\frac{2}{ln2}$ to double its sending rate to probe more available bandwidth. When the packet received rate is 1.25 times less than the previous sending rate and lasts for 3 times, the sender deems it approach to the available bandwidth and enters into the Drain phase. The Drain phase is applied to decrease the sending rate to get rid of excess queue accumulated during the StartUp phase. When the inflight packets are less than the computed BDP (Bandwidth Delay Product), the state is changed to ProbeBW phase. At this state, the sending rate is controlled by different gain value [1.25, 0.75, 1, 1, 1, 1, 1, 1] in 8 RTT periods. The gain value 1.25 is exploited for bandwidth probe purpose. If there is extra bandwidth available, the sender increases its rate to occupy the extra bandwidth. During a window of 10 seconds, if the minimal RTT signal is not detected, the link is deemed in congestion status and the state will be transferred to ProbeRTT and only four packets are sent out. The ProbeRTT state will last at most 200 milliseconds, which can be seen as a synchronization operation to give the new comer flows an opportunity to probe their fair bandwidth share.

The original BBR algorithm is not appropriately applied for real time video transmission. First, the endpoint would send to many packets during the ProbeBW phase with the pacing gain 1.25 to probe bandwidth. At the next RTT, the transmission rate would be reduced to 0.75 times than its available bandwidth. This sharp rate reduction would make the video packets queued at the local buffer and extra delay thus is introduced. Second, when the BBR flows competing bandwidth, all the flows would insist the maximum rate sampled during the last 10 RTT periods and the total sending rate would exceed the link capacity. The excess sent packets would be queued at the intermedia routers and the transmission delay would increase. The minimal RTT would not be sampled during 10 seconds (kMinRttExpiry) monitoring interval. And the ProbeRTT phase which lasts at most 200 milliseconds may be not long enough to let the intermediate routers drain the occupied buffer. Thus the optimal control point is deviated. Hence, high packet loss rate and transmission delay are observed in experiments. Third, only four packets are sent during ProbeRTT phase would further introduce queue delay at the sender buffer.

Given the above mentioned drawbacks of the BBR algorithm, an optimized version for real time video transmission is proposed in this work. First, the pacing gain values [1.25, 0.75, 1, 1, 1, 1, 1, 1] are changed to [1.11, 0.9, 1, 1, 1, 1, 1, 1] in 8 RTT periods during the ProbeBW phase in order to keep the packet sending rate stable. The 1.25X pacing gain shows quite aggressiveness when there is extra bandwidth available when flows competing for resource. Second, as for optimal control point deviation problem due to bandwidth scramble, instead of only entering the ProbeRTT state when the minimal RTT is not sampled in 10 seconds, we found that let the sender enter the ProbeRTT state will make the intermediate routers to drain the queued packet and improve the performance of the congestion control algorithm. Hence, the BBR algorithm is enhanced to respond the link delay signal by actively reduce sending rate to let the occupied buffer drain and make the algorithm to get close to its optimal control point. The smoothed RTT (SRTT) signal is monitored during the ProbeBW phase and is computed as Equation \eqref{eq:srtt}. $\alpha$ is 0.9 and the exponential filter is applied to eliminate delay signal noise. Once srtt is $\beta$ times larger than the base RTT (base\_line\_rtt\_, the minimal RTT during the ProbeBW phase), the link seems fall into congestion (line 4 in Algorithm \ref{alg:checkifcongestion}), and the control state would be transferred into ProbeRTT. Here, $\beta$ is set as 1.2. And rate pacing gain will be set as 0.75, which means the sending rate would be reduced to $0.75X$ of its maximum rate instead of only four packets are sent in a RTT period. Until the inflight packets are less than the BDP (bw*min\_rtt\_), the state would be changed into ProbeBW (line 11 in Algorithm \ref{alg:enterdecrease}). In such case, the purpose of ProbeRTT is not exploited for sampling a new minimal RTT but to drain the excess sent packets, which is different to its original version. The state would be changed to ProbeRTT for many times during the 10 seconds of the minimal RTT monitoring interval in the modified algorithm. Once the control state is set as ProbeRTT due to increased latency, srtt would be reset as zero (line 4 in Algorithm \ref{alg:enterdecrease}) until the acknowledged sequence number is larger than the packet number (seq\_at\_backoff\_, line 3 in Algorithm \ref{alg:enterdecrease}) sent before the state ProbeRTT. Such operation is to make the algorithm more robust when the flow sharing links with loss base congestion control constraint flows. The optimized version BBR algorithm for real time video transmission is named as delay response BBR (Delay-BBR).
\begin{equation}
\label{eq:srtt}
srtt=(1-\alpha)*srtt+\alpha*rtt
\end{equation}
\begin{algorithm}[htb] 
\caption{OnPacketSent} 
\label{alg:onsent} 
\begin{algorithmic}[1] 
\REQUIRE ~~\\ 
the timestamp (now), sequence (seq) and the payload length of a sent packet
\STATE info.sent\_ts=timestamp,info.bytes=payload
\STATE sent\_packets\_map\_.insert(seq,info)
\STATE inflight\_$\gets$ inflight\_+payload
\STATE last\_sent\_packet\_$\gets$seq
\end{algorithmic}
\end{algorithm}
\begin{equation}
\label{eq:bw}
bw=min(\frac{\Delta sent}{\Delta sent\_ts},\frac{\Delta acked}{\Delta ack\_ts})
\end{equation}
\begin{algorithm}[htb] 
\caption{OnAck} 
\label{alg:onack} 
\begin{algorithmic}[1] 
\REQUIRE ~~\\ 
the ack packet received timestamp (now) and acknowledged sequence number (seq)
\STATE UpdateRttAndInflight(now,seq)
\STATE congested$\gets$CheckIfCongestion()
\IF{now-min\_rtt\_ts\_$>$kMinRttExpiry}
\STATE min\_rtt\_expired$\gets 1$
\ENDIF
\STATE MaybeEnterOrExitDrain(now,min\_rtt\_expired,congested)
\end{algorithmic}
\end{algorithm}
\begin{algorithm}[htb] 
\caption{UpdateRttAndInflight} 
\label{alg:updatertt} 
\begin{algorithmic}[1] 
\REQUIRE ~~\\ 
the ack packet received timestamp (now) and acknowledged sequence number (seq)
\STATE get the sent packet info in sent\_packets\_map\_
\STATE rtt=info.sent\_ts$-$now
\STATE inflight\_$\gets$ inflight\_$-$info.bytes 
\IF{rtt$<$min\_rtt\_ \OR min\_rtt$==$0}
\STATE min\_rtt\_$\gets$rtt
\STATE min\_rtt\_ts\_$\gets$now
\ENDIF
\IF{rtt$<$kSimilarMinRtt*min\_rtt\_}
\STATE min\_rtt\_ts\_$\gets$now
\ENDIF
\IF{seq$>$seq\_at\_backoff\_}
\IF{rtt$<$base\_line\_rtt\_}
\STATE base\_line\_rtt\_$\gets$rtt
\STATE srtt\_$\gets$rtt
\ENDIF
\STATE srtt\_$\gets(1-\alpha)$*srtt\_+$\alpha$*rtt
\ENDIF
\end{algorithmic}
\end{algorithm}
\begin{algorithm}[htb] 
\caption{CheckIfCongestion} 
\label{alg:checkifcongestion} 
\begin{algorithmic}[1] 
\IF{srtt\_$==$0 \OR base\_line\_rtt\_$==+\infty$}
\RETURN $0$
\ENDIF
\IF{mode\_$==$ProbeBW \AND srtt\_$>\beta$*base\_line\_rtt\_}
\RETURN $1$
\ENDIF
\end{algorithmic}
\end{algorithm}
\begin{algorithm}[htb] 
\caption{MaybeEnterOrExitDrain} 
\label{alg:enterdecrease} 
\begin{algorithmic}[1] 
\REQUIRE ~~\\ 
(now, min\_rtt\_expired, congested)
\IF{mode\_$!=$ProbeRTT \AND (min\_rtt\_expired \OR congested)}
\STATE mode\_$\gets$ProbeRTT
\STATE seq\_at\_backoff\_$\gets$last\_sent\_packet\_
\STATE srtt\_$\gets$0
\STATE base\_line\_rtt\_$\gets+\infty$
\STATE pacing\_gain\_$\gets 0.75$
\STATE bdp\_=bw\_*min\_rtt\_
\ENDIF
\IF{mode\_$==$ProbeRTT}
\IF{inflight\_$<$ bdp\_}
\STATE EnterProbeBwMode()
\ENDIF
\ENDIF
\end{algorithmic}
\end{algorithm}
At the sender sider, on each sent packet, the packet length and sent timestamp will be recorded as shown in Algorithm \ref{alg:onsent}. Here, inflight\_ is the total sent packets length without being acknowledged. On each acknowledged packet that reports the sent packets received information, the sender would compute the bandwidth as Equation \eqref{eq:bw} and follow the Algorithm \ref{alg:onack} to update the rtt information and change the packet inflight counter in Algorithm \ref{alg:updatertt}, check whether the link is in congestion status in Algorithm \ref{alg:checkifcongestion}. When the link delay increase or the minimal rtt timestamp expires (kMinRttExpiry is 10 seconds), the rate pacing gain will be set as 0.75 in Algorithm \ref{alg:enterdecrease} to let the buffered packets drain and to maintain lower transmission latency.

Even though I have tried my best to explain the algorithm detail both in word and pseudo code, the weakness of the language would make the algorithm implementation hard for interested readers, thus the code is released at github \footnote{https://github.com/SoonyangZhang/congestion-responce-experiment}.

One more thing should be put forward is that when the proposed algorithm applied for real time video transmission, padding packets which carry useless data must be generated for bandwidth probe purpose when there are no video packets available.
\subsection{The proposed packet scheduling algorithm}
The proposed multipath schedule algorithm is working on packet level. When a video frame comes, the packed packets would be sent to the schedule module to choose the path with the minimal path cost. And the cost of path $p$ in the proposed algorithm is defined as the following:
\begin{equation}
\label{eq:cost}
\lambda_p=\frac{RTT_p(t) }{2}+\frac{Q_p(t)}{c_p(t)}
\end{equation}

$RTT_p(t)$ is round trip time. This value of $\frac{RTT_p(t)}{2}$ can be used as a reflection of the link queue delay and propagation delay. $Q_p(t)$ is the length of the pending packets waiting to be transmitted out by the Pacer. When the packets of a new frame are put to the sent buffer to path P, the sent buffer may not be empty and they will be buffered. There will be a waiting delay before these packets can be scheduled into network link by the pacer. $c_p(t)$ is the packet sending rate of the Pacer and the rate is determined by the congestion control module. On each incoming packet, the schedule module will choose the path with the minimal path cost as indicated by equation \eqref{eq:min-schedule}. The detail schedule procedure is shown in Algorithm \ref{alg:schedule}.
\begin{equation}
\label{eq:min-schedule}
j=\arg\min_p{\lambda_p}
\end{equation}
\begin{algorithm}[htb] 
\caption{Packets schedule Procedure} 
\label{alg:schedule} 
\begin{algorithmic}[1] 
\REQUIRE ~~\\ 
The id and length of newly generated packets
\ENSURE ~~\\ 
the map between packet id and path id
\REPEAT 
\STATE Compute each path cost $\lambda_p$ as equation \eqref{eq:cost}
\STATE Get the path index $j$ with the minimal path cost
\STATE Schedule packet $i$ to path $j$
\STATE Increase the buffer pending length $Q_j$ of path $j$
\UNTIL{all incoming packets are scheduled}
\end{algorithmic}
\end{algorithm}

From the definition of path cost function, to choose the path with minimal cost would deliver the packet to the receiver with the minimal delay. Even it cannot strictly guarantee in order of delivery when packets arrive to the destination due to some uncertain factors in the routing path, e.g., the exact buffered packets length in the intermediate routers, the scheduling algorithm strives to achieve a minimized arriving interval between two consecutive packets. A minimal arriving interval between packet could make the received packets can be recombined as video frame and to be delivered to the upper layer as early as possible.

In single transmission scheme, a receiver buffer is needed to counter packets arriving jitter and waiting for the packets can be recombined as completing frame. In the proposed multipath framework, the packet receive buffer is working on session level as shown in Figure \ref{Fig:system}, which is different from the path level buffer at the sender side. Due to the packets belonging to the same frame can be scheduled to different paths. And the receiver will not exactly know an uncompleted frame is caused by packet lost or packet late. Hence, at the path level, each sub-path would feedback the packets received information to the sender and whether to retransmit the lost packets will be totally decided by the sender. The sent packets will be kept at the sender session level buffer at most 500 milliseconds. If the lost packet is found in the sender session level buffer, it will be retransmitted immediately to the path with minimal transfer delay. The receiver will decide whether to wait for the lost packet in current uncompleted frame, since the retransmission process will introduce extra delay. For key frame, receiver must wait the lost packet to be retransmitted and deliver a complete frame, or else the following frames will not be decoded successfully, which has a severe impact on QoE. For the other frames, if lost packet event happens, the incomplete frame will be dropped if the waiting time exceeds a maximum waiting time threshold (500 milliseconds in experiment) set by the user.
\subsection{Implementation of benchmark algorithm}
Three other packets scheduling algorithms Weight Round Robin (WRR), EDCLD \cite{Prabhavat2011Effective}, and SFL \cite{Wu2015low} are implemented in current simulation framework, for performance comparison purpose. 

The WRR algorithm is originating from server load balance. The weight of each path takes the form in equation \eqref{eq:wrr-schedule}. The packets are scheduled based on path available bandwidth.
\begin{equation}
\label{eq:wrr-schedule}
path\_load(i)=\lceil n\frac{ABW_i}{\sum_{j}ABW_j}\rceil
\end{equation}
\begin{algorithm}[htb] 
\caption{Water filling algorithm} 
\label{alg:sfl-schedule} 
\begin{algorithmic}[1] 
\REQUIRE ~~\\ 
The $path\_info(ABW, OWD)$ every available path,\\
$n$, total available path,\\
The total size $s$ and $packet\_info(id,len)$ of newly generated packets
\ENSURE ~~\\ 
the map of  packet id and path id
\STATE sort the $path\_info$ according to owe way delay
\STATE $L\gets packet\_info[start\_id].len$
\STATE $step\gets \frac{L*1000*8}{path\_info[n-1].ABW}$
\FOR{$i \in [0,\infty)$}
\STATE $compensator\gets i*step$
\STATE $D\gets path\_info[n-1].OWD+compensator$
\STATE $water\gets 0$
\FOR{$i \in [0,n)$}
\STATE $owd\gets path\_info[i].OWD$
\STATE $abw\gets path\_info[i].ABW$
\STATE $path\_info[i].bytes=\frac{(D-owd)*abw}{1000*8}$
\STATE $water\gets water+path\_info[i].bytes$
\ENDFOR
\IF{$water\geq s$}
\STATE \bf{break}
\ENDIF
\ENDFOR
\STATE ${i\gets 0}$
\FOR{$k \in [start\_id,end\_id]$}
\WHILE{ $i\neq n-1$\AND$path\_info[i].bytes<0$}
\STATE $i\gets i+1$
\ENDWHILE
\STATE Schedule packet $k$ to path $i$
\STATE $len=packet\_info[start\_id].len$
\STATE $path\_info[i].bytes\gets path\_info[i].bytes-len$
\ENDFOR
\end{algorithmic}
\end{algorithm}
\begin{equation}
\label{eq:edcld-cost}
C_p(\psi_p)=D_p+(1-w)\frac{1}{\mu_p-\psi_p\lambda}+w\frac{q}{\mu_p}
\end{equation}
\begin{figure}
\centering
\includegraphics[width=3in]{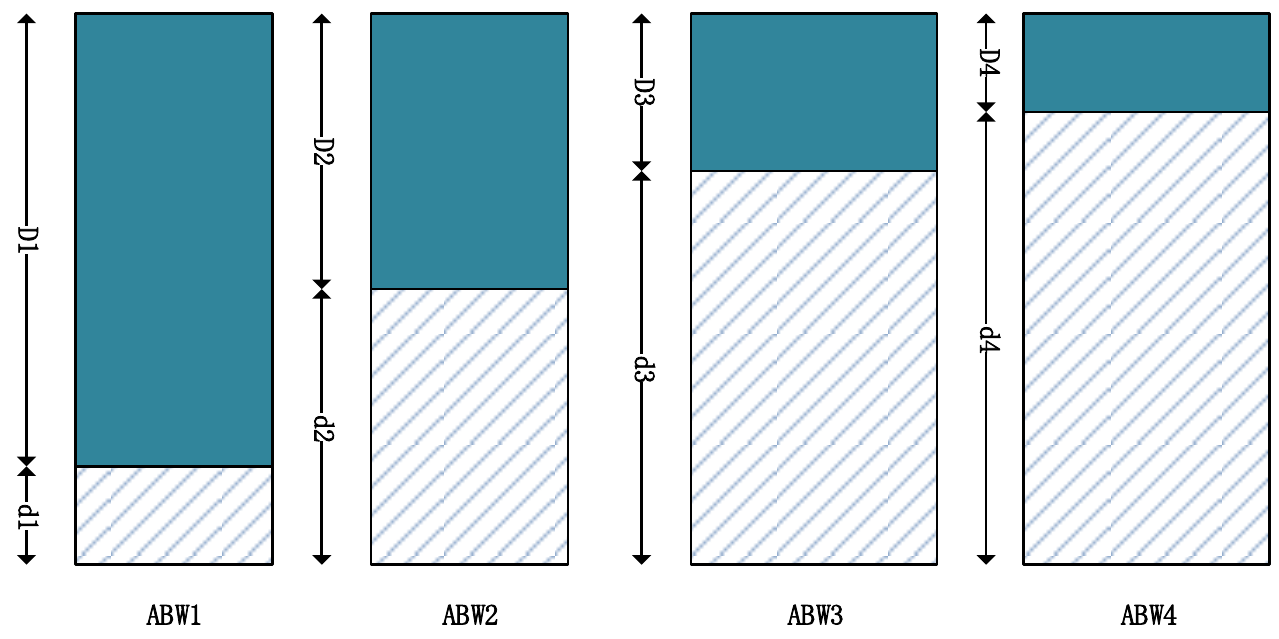}
\caption{watering filling method}
\label{Fig:water}
\end{figure}

The path cost function defined in EDCLD shown in equation \eqref{eq:edcld-cost} is a hybrid consideration of Poission traffic and unknown traffic. Here, $D_p$ is the fix propagation delay, $\mu_p$ is the available bandwidth, $q_p$ is the input buffer,$psi$ is the packet splitting ratio, $w$ is the weight and the cost function follows M/M/1 queue model when $w\gets 0$. In experiment, $w$ is set to 0.8. It will increase the packets splitting ratio on the best path with lowest cost value and decrease it on the worse path. In order to find an appropriate packet splitting ratio, a quadratic equation $C_{p_{best}}(\psi_{best}+\Delta\psi)=C_{p_{worst}}(\psi_{worst}-\Delta\psi)$ needs to be solved. In an iteration, the packets splitting ratio on the paths with the cost between $C_{p_{best}}$ and $C_{p_{worst}}$ will remain unchanged.

The water filling algorithm implemented in SFL takes link available bandwidth and delay into consideration. The basic idea behind water filling algorithm is to make an equal aggregate delay through scheduling more packets to the path with smaller transmission delay and less packets to larger transmission delay path as shown in Figure \ref{Fig:water}, namely $D_i+d_i=D_j+d_j$. Here, $d_i$ is the one way delay of path $i$, $D_i$ is the water depth. Through the adjustment of $D_i$, the total packets that all paths can be sent in a schedule round should be equal to the incoming packets, $\sum_iD_i*ABW_i=length$. The detail of the water filling packet algorithm is shown in Algorithm \ref{alg:sfl-schedule}. In order to simplify the calculation process, when the current water of all path cannot hold the incoming packets, the water depth of the path with the largest one way delay will increase by a step of $\Delta D=\frac{L}{ABW}$ until all the packet can be sent out, and $L$ is the length of a meta packet.
\section{Evaluation}
All the simulation experiments are conducted in ns-3.26 on Ubuntu 14.04 in VMware.
\subsection{Performance of the congestion control algorithm}
In order test the performance of the optimized version of BBR algorithm, a point to point channel is built with link configuration shown in Table \ref{tab:link-conf}. To make a comparison, we get the BBR implementation in google QUIC codebase running on ns3 platform. The experiments were running about 300 seconds. 
\begin{table}[]
\centering
\caption{Network configuration}
\label{tab:link-conf}
\begin{tabular}{|c|c|c|c|}
\hline
Case & Bandwidth& One Way Delay& Queue buffer \\ \hline
1 & 3Mbps & 100ms & 3Mbps*300ms \\ \hline
2 & 3Mbps & 100ms & 3Mbps*400ms \\ \hline
3 & 3Mbps & 100ms & 3Mbps*600ms \\ \hline
4 & 4Mbps & 100ms & 4Mbps*300ms \\ \hline
5 & 4Mbps & 100ms & 4Mbps*400ms \\ \hline
6 & 4Mbps & 100ms & 4Mbps*600ms \\ \hline
7 & 5Mbps & 100ms & 5Mbps*300ms \\ \hline
8 & 5Mbps & 100ms & 5Mbps*400ms \\ \hline
9 & 5Mbps & 100ms&  5Mbps*600ms \\ \hline
\end{tabular}
\end{table}

\begin{figure}
\centering
\includegraphics[width=3in]{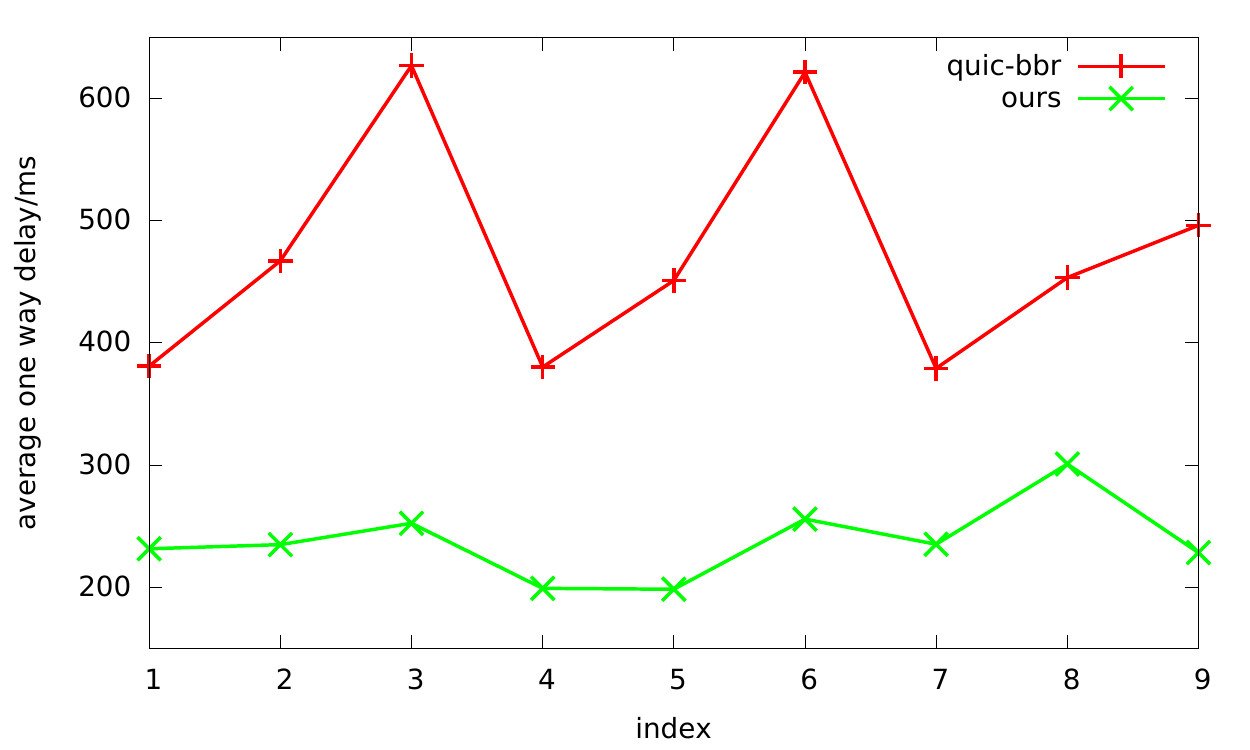}
\caption{Average packet owe way transmission delay}
\label{Fig:owd-comapre}
\end{figure}

\begin{figure}
\centering
\includegraphics[width=3in]{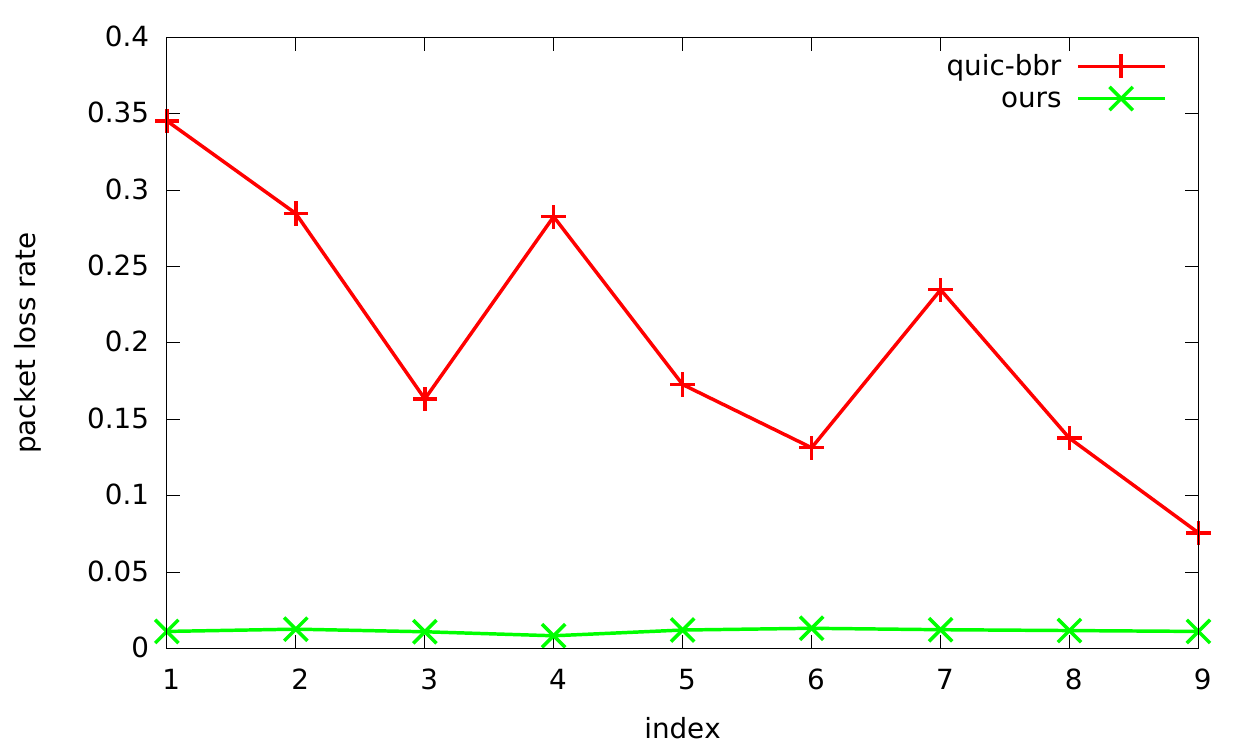}
\caption{Average packet loss rate}
\label{Fig:loss-comapre}
\end{figure}

On each experiment, three flows exploiting the same congestion control algorithm are initiated at different time. The second flow was started after 40s later than the first flow and the third flow was started at 80s. At the sender side, the packet sending rate is traced. Packet one way transmission delay and the loss packet sequence number are recorded at the receiver side. The average one way transmission delay and average packet loss rate are calculated of all three flows. The average one way transmission delay of each test case is plotted in Figure \ref{Fig:owd-comapre} And the packet loss rate of all 9 experiments is showed in Figure \ref{Fig:loss-comapre}.

The packet one way transmission delay indicates whether excess packets are sent into the network by network users and much buffer resource is occupied. Our optimized BBR algorithm achieves lower packet transmission and extremely lower packet loss rate compared with the QUIC-BBR. As the buffer length increase (Experiment 1, 2, 3), the packet loss rate is decrease while transmission delay is increase in QUIC-BBR. The bandwidth aggressiveness property in QUIC-BBR would occupy much buffer resource.

\begin{figure}
\centering
\includegraphics[width=3in]{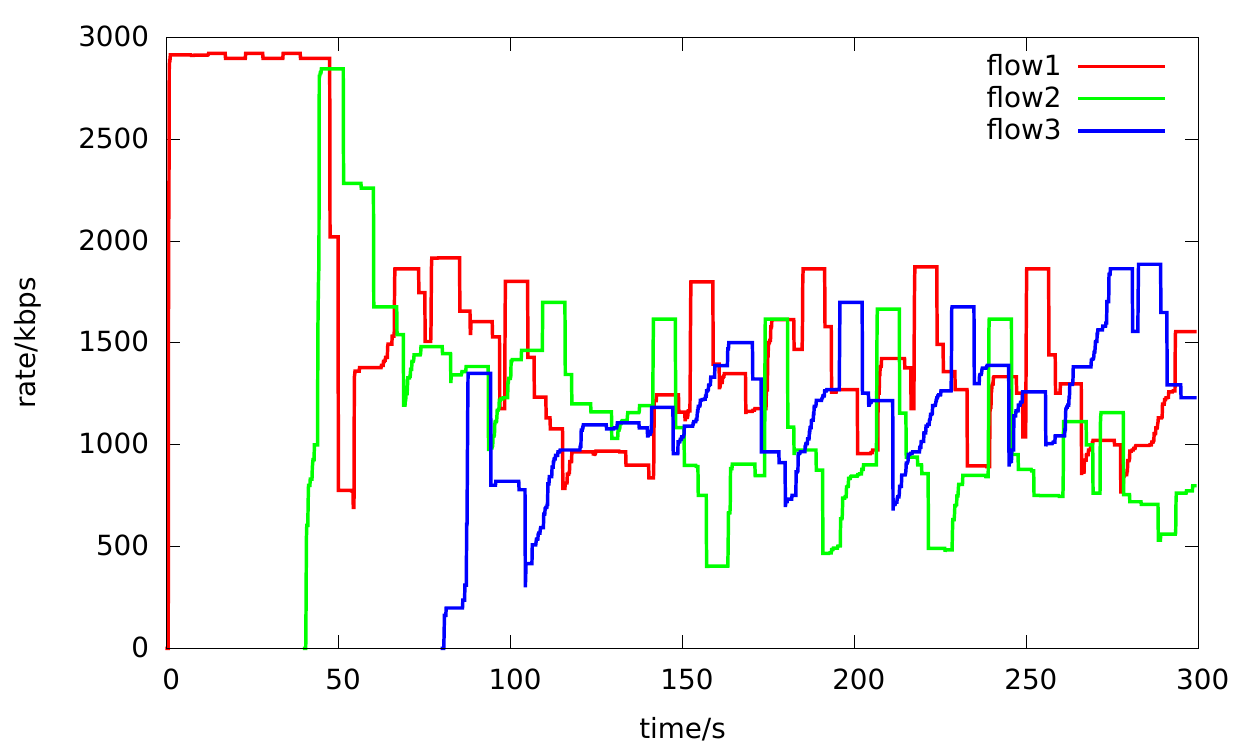}
\caption{Sending rates of  QUIC-BBR flows}
\label{Fig:quic-bbr-dynamic}
\end{figure}

\begin{figure}
\centering
\includegraphics[width=3in]{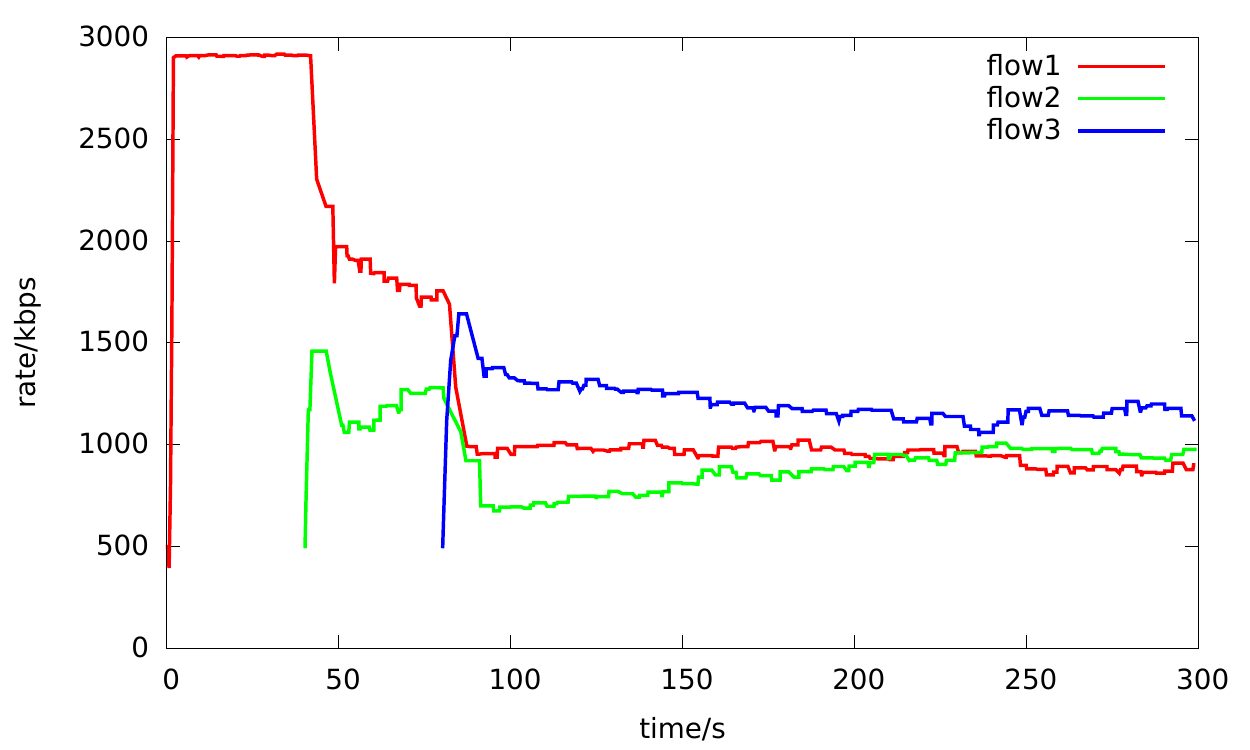}
\caption{Sending rates of Delay-BBR flows}
\label{Fig:mybbr-dynamic}
\end{figure}

\begin{figure*}
\centering
\subfigure[E1]{\includegraphics[width=2in]{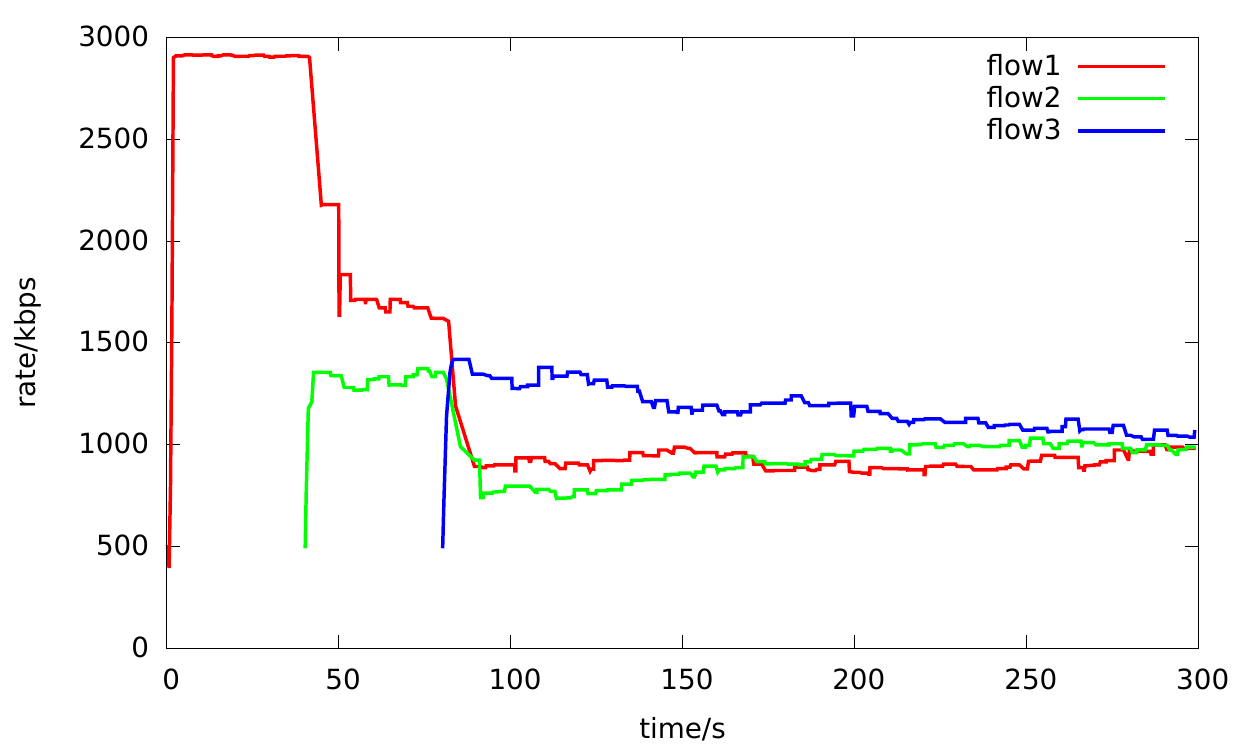}}
\subfigure[E4]{\includegraphics[width=2in]{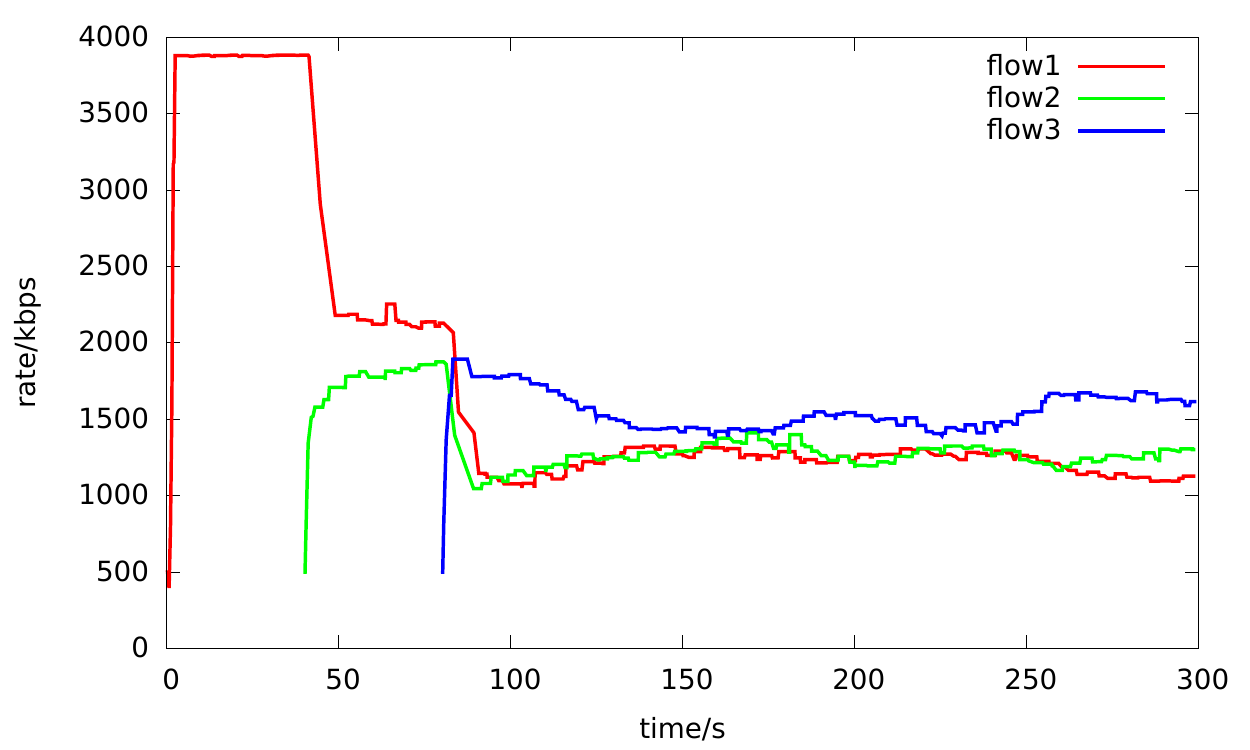}}
\subfigure[E7]{\includegraphics[width=2in]{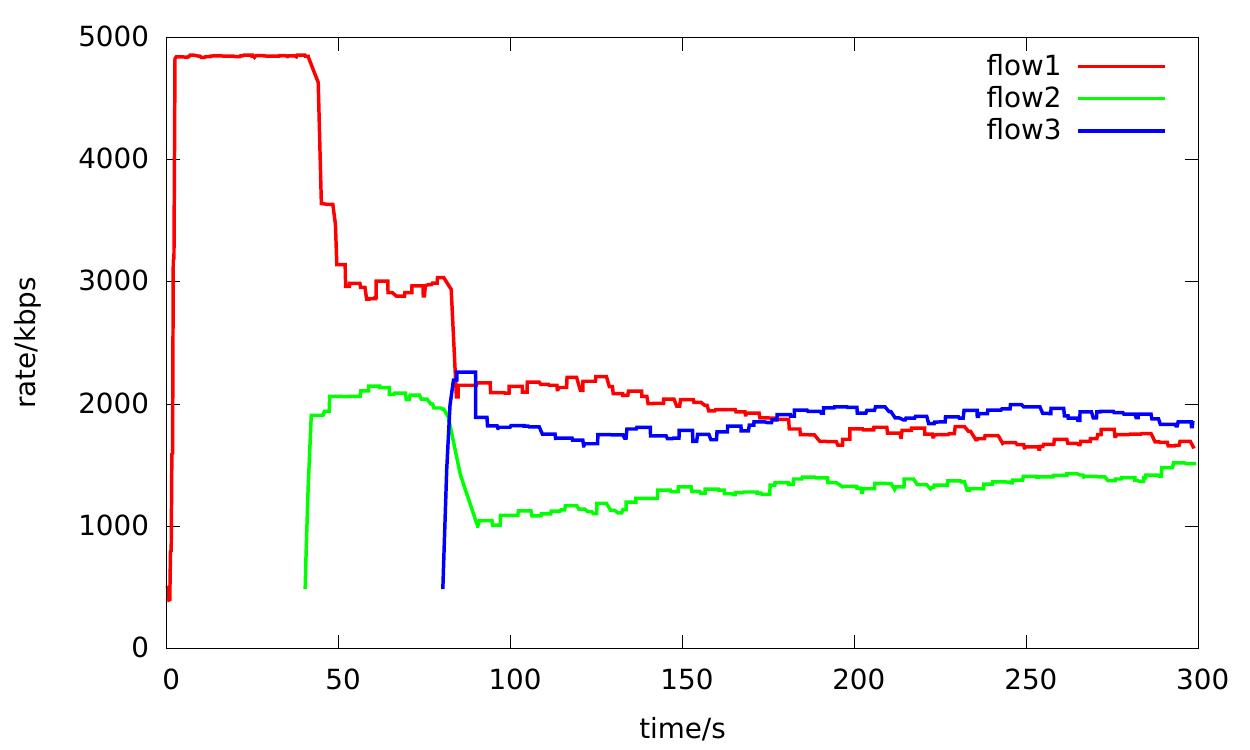}}
\caption{Sending rates of Delay-BBR flows}
\label{Fig:mybbr-1-4-7}
\end{figure*}

The sending rates of QUIC-BBR flows and optimized BBR flows are plotted in Figure \ref{Fig:quic-bbr-dynamic} and Figure \ref{Fig:mybbr-dynamic} under the link configuration 2, in which  the link capacity is 3Mbps. The rates of QUIC-BBR flows can maintain fairness on average but show a wide range of variation from 500kbps to 1500bps due to the bandwidth competition. While the rates of the optimized version BBR flows show stability and converge to fairness line near 1Mbps Figure \ref{Fig:mybbr-dynamic}. A single test case may be not quite persuasive, Figure \ref{Fig:mybbr-1-4-7} further shows such rate convergence property, or in other word, the optimized algorithm can maintain bandwidth allocation fairness. The rate stability property of congestion control algorithm make it fit for video transmission.

The reason for high packet loss rate and transmission delay is that all the flows tend to insist the maximum computed bandwidth accord to the packets acknowledged rate and the total sending rate would exceed the link capacity as shown in Figure \ref{Fig:quic-bbr-dynamic}. In such situation, the occupied link buffer cannot be drained in time and packet transmission delay increases. When the occupied link buffer overflows, packet loss will happen. In different to the implementation of QUIC-BBR, we make the modified algorithm response delay signal as detailed in algorithm implementation part. When the link delay has exceeded the defined threshold, the sending rate will be reduced to let the intermediate routers to drain the excess occupied buffer, thus a lower delay can achieve in our implementation.

\begin{figure}
\centering
\includegraphics[width=3in]{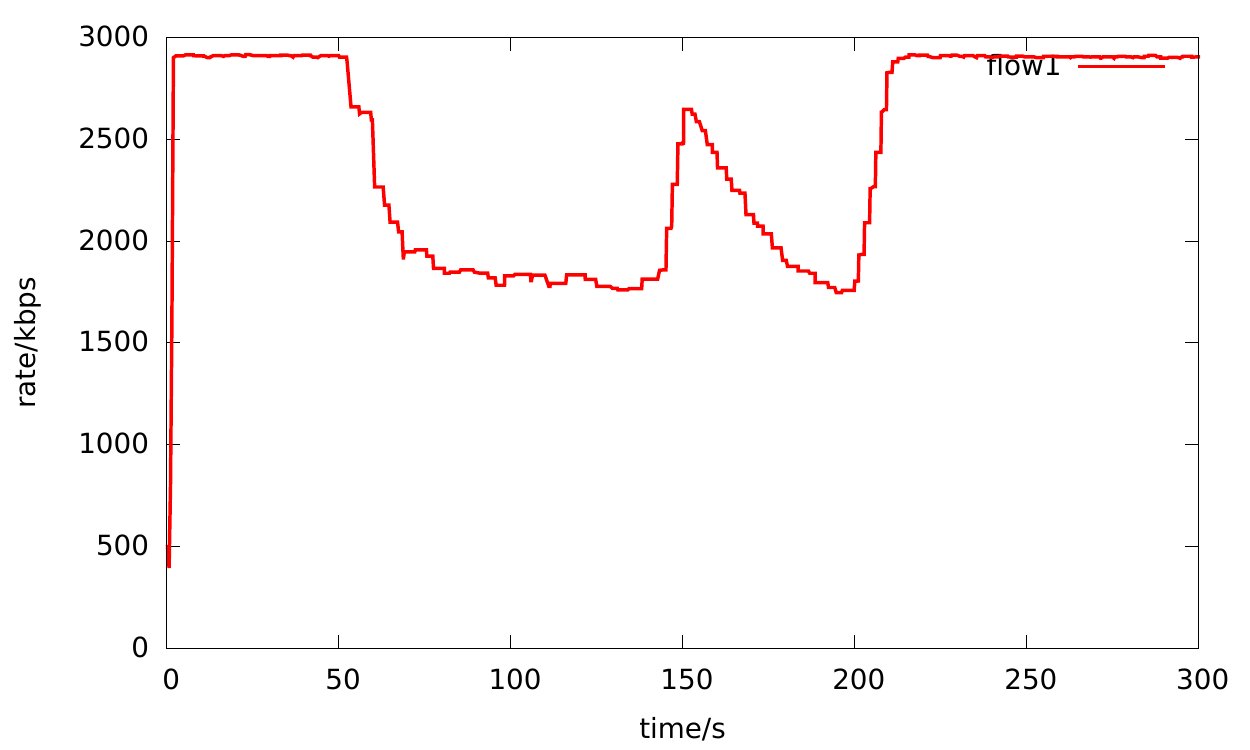}
\caption{Sending rate of Delay-BBR flow}
\label{Fig:mybbr-tcp-dynamic}
\end{figure}

Traditionally, the delay response congestion control algorithms such as Vegas tend to get starvation when sharing links with loss based congestion control algorithms. The proposed algorithm does not suffer from the bandwidth starvation problem. An experiment is conducted that a TCP Reno flow and an optimized BBR flow shared link in case 2 configuration to support such claim. The TCP Reno flow is started at 50 seconds and ended at 200 seconds. The flow can maintain reasonable high throughput and indeed yields bandwidth during the presence of TCP Reno flow in Figure \ref{Fig:mybbr-tcp-dynamic}. After the TCP Reno flow stops, the flow rate rapidly increases to a point close to the link capacity. Thanks to the reset operation of srtt signal when the state is changed to ProbeRTT to let the queue buffer drain and srtt would be resampled in state ProbeBw. Such operation can prevent the control state be changed to ProbeRTT too frequently to yield too much bandwidth.

\begin{figure*}
\centering
\subfigure[E1]{\includegraphics[width=2in]{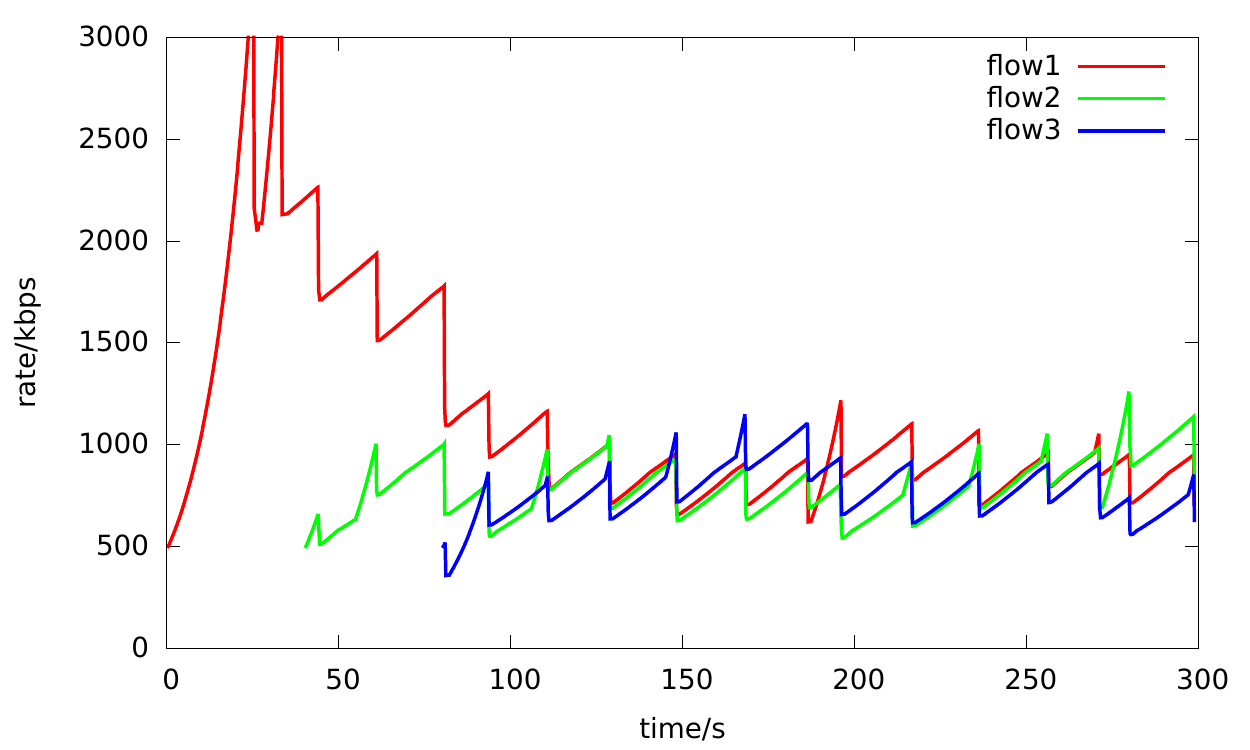}}
\subfigure[E4]{\includegraphics[width=2in]{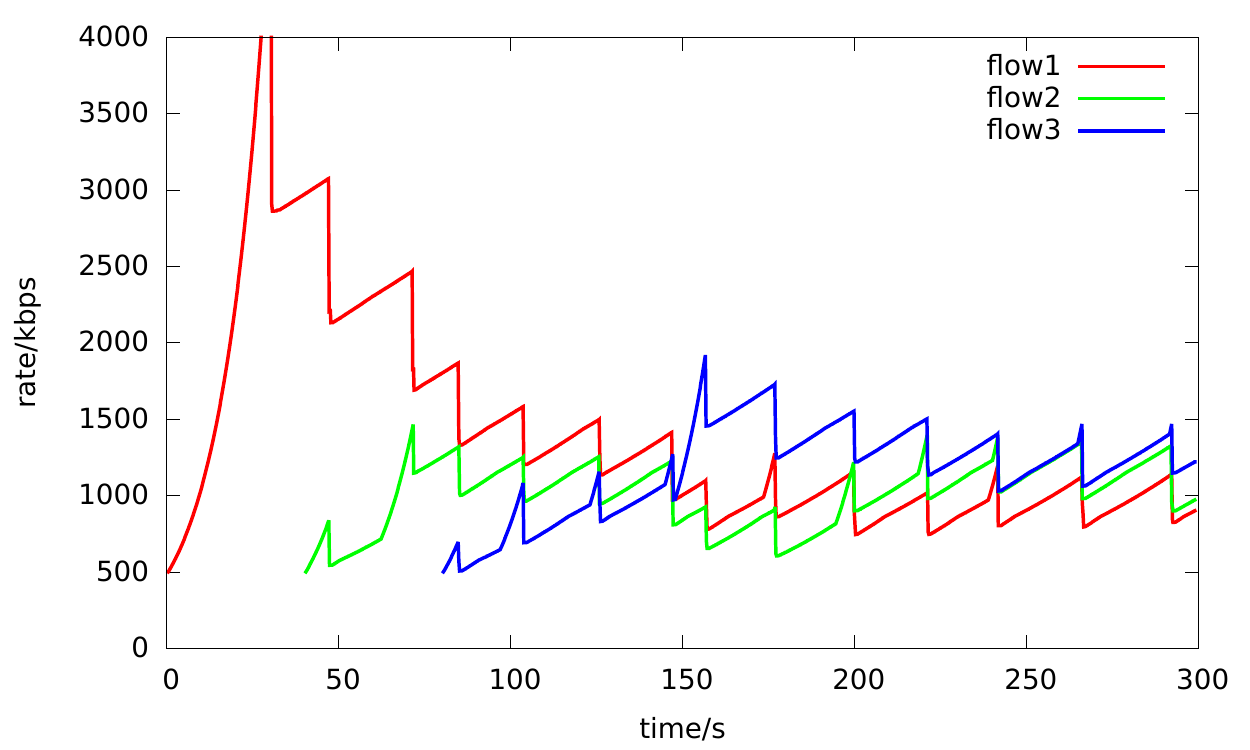}}
\subfigure[E7]{\includegraphics[width=2in]{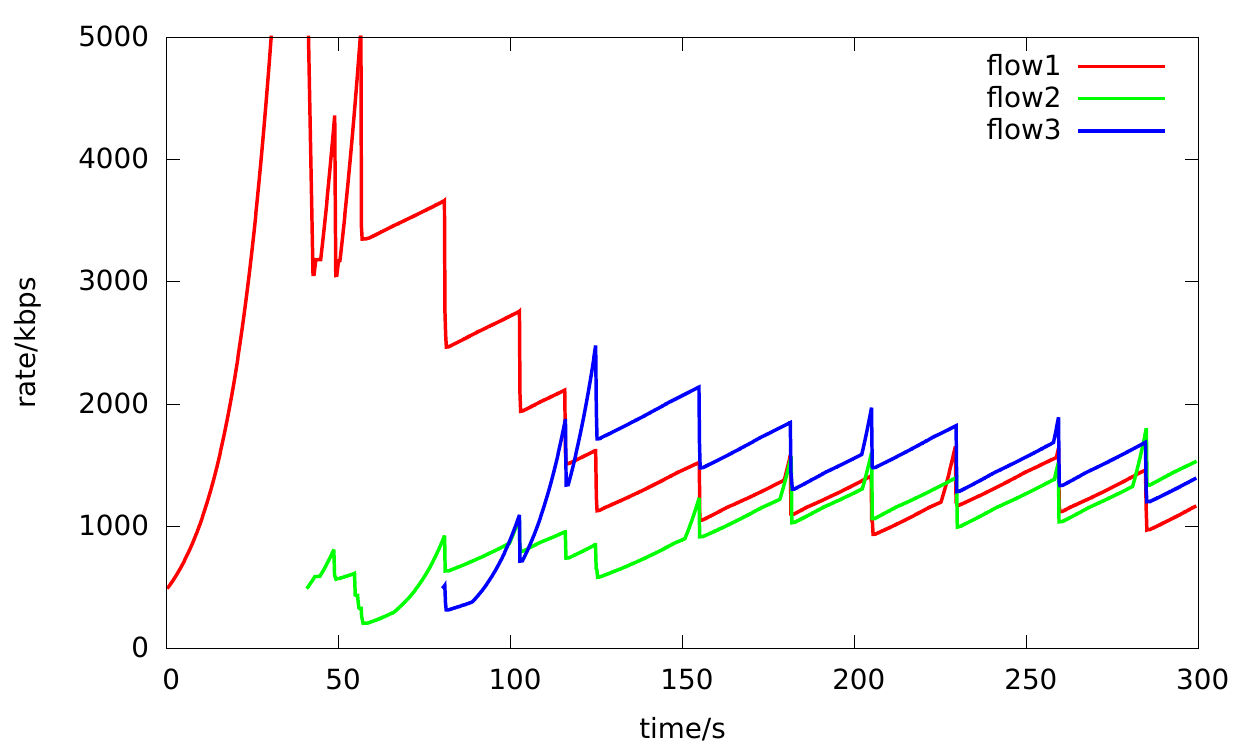}}
\caption{Sending rates of WebRTC-GCC flows}
\label{Fig:webrtc-gcc-1-4-7}
\end{figure*}

\begin{figure}
\centering
\subfigure[E1]{\includegraphics[width=2in]{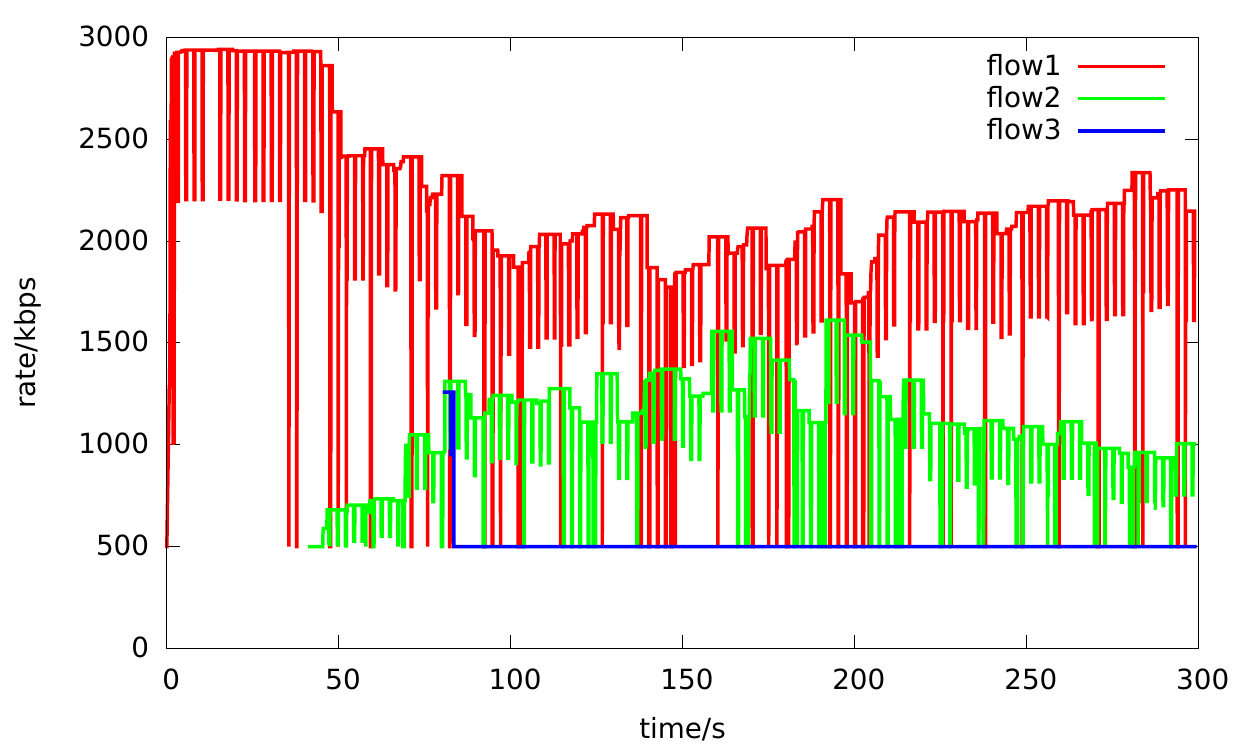}}
\subfigure[E4]{\includegraphics[width=2in]{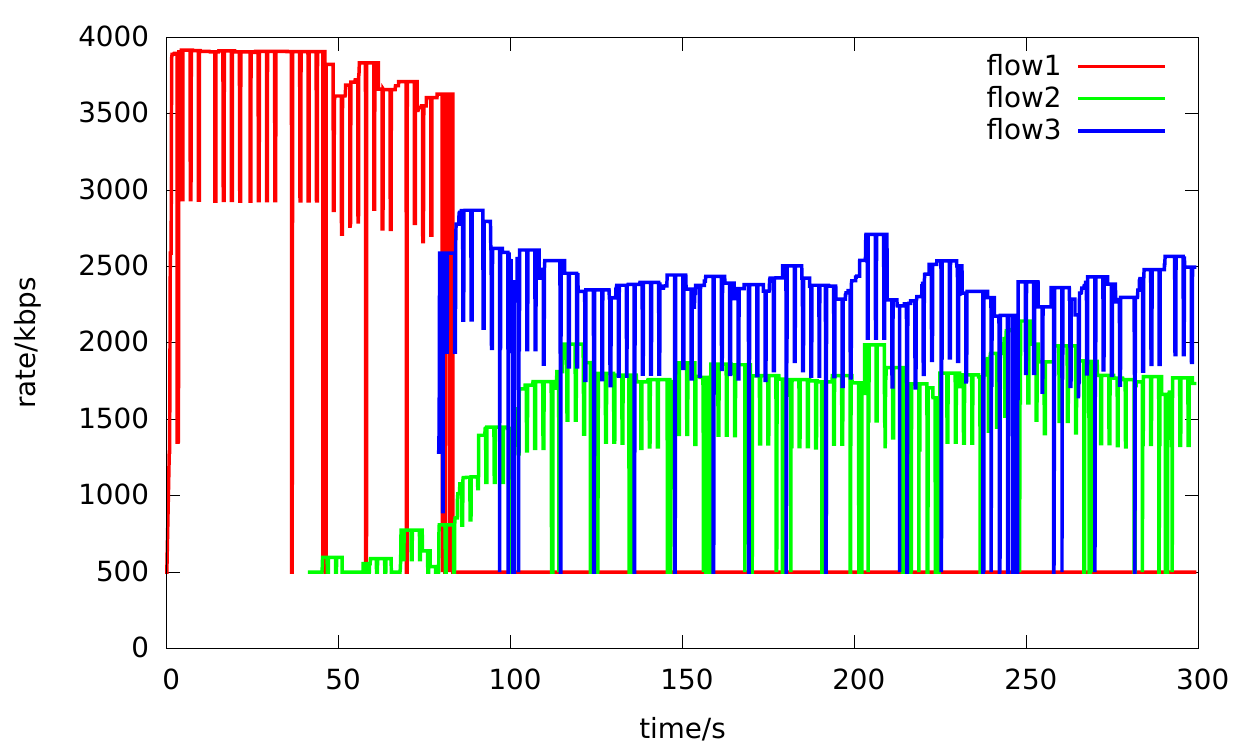}}
\subfigure[E7]{\includegraphics[width=2in]{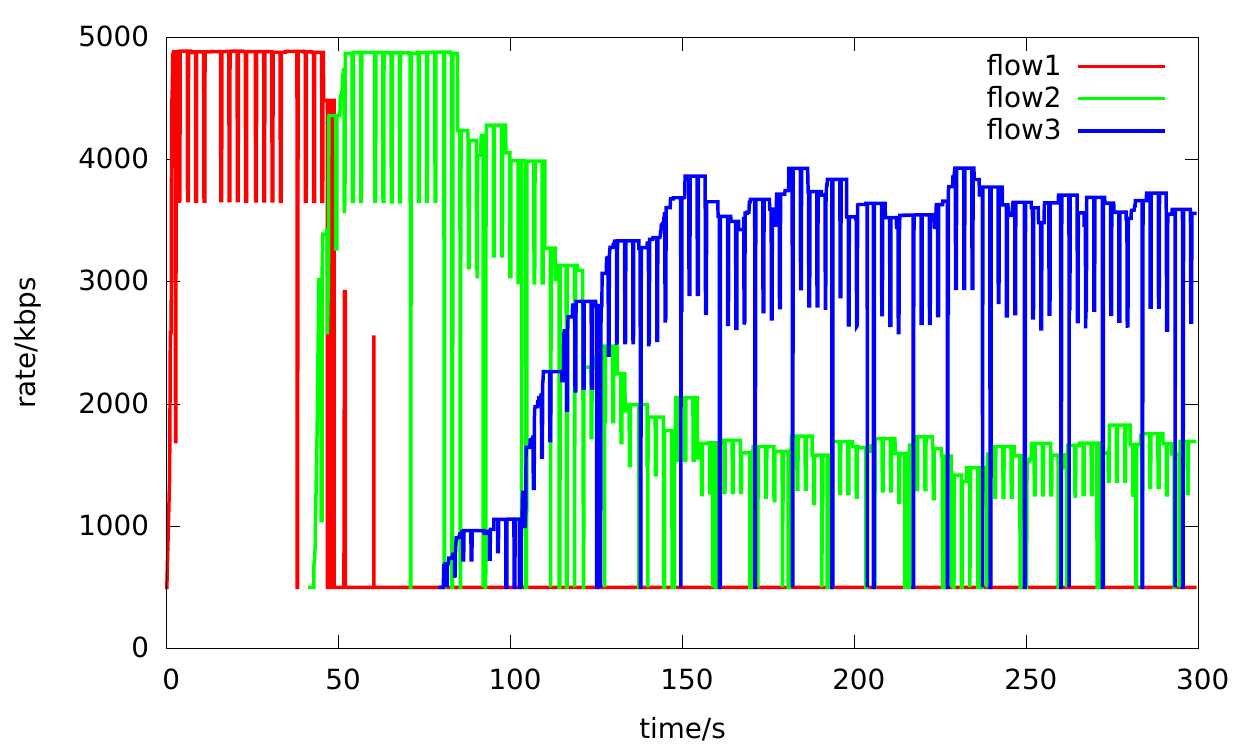}}
\caption{Sending rates of WebRTC-BBR flows}
\label{Fig:webrtc-bbr-1-4-7}
\end{figure}

Currently, the default applied congestion control algorithm in WebRTC is GCC. In an older version, the BBR algorithm was implemented in WebRTC codebase, but in newer version, the part on BBR is not compiled, and there is not any report on the performance of WebRTC-BBR in real network.  In our previous work \cite{Zhang2018Congestion}, the performance comparison on GCC, NADA, SCReAM and WebRTC-BBR is conducted in ns3 platform. Here, the performance comparison of GCC, WebRTC-BBR and our optimized version of BBR is tested with the same link configuration in case 1, 4,7. In each case, three flows are tested and the experiment are running about 300s. And the final results on sending rates of all three flows with different congestion control algorithm are shown in Figure \ref{Fig:mybbr-1-4-7}, Figure \ref{Fig:webrtc-gcc-1-4-7}, Figure \ref{Fig:webrtc-bbr-1-4-7}. From Figure \ref{Fig:webrtc-gcc-1-4-7} and Figure \ref{Fig:mybbr-1-4-7} , both GCC flows and optimized BBR flows converging to fairness bandwidth line shown can be concluded. While the rates of the WebRTC-BBR flows show oscillation and a flow cannot get the fairness bandwidth and only a minimal rate is maintained due to other flows occupy much bandwidth resource as Figure \ref{Fig:webrtc-bbr-1-4-7} shows.

Further, the average packet transmission delay and packet loss rate is shown in Table \ref{Tab:algorithm-delay-loss}. In the same link configuration, GCC flows can maintain the lowest transmission delay but have considerable packets loss rate about 10\%. The optimized BBR flows have the lowest packet loss rate which is almost negligible but have higher average packet transmission delay compared with GCC.
\begin{table}[]
\centering
\caption{Transmission delay and loss rate of compared algorithms}
\label{Tab:algorithm-delay-loss}
\begin{tabular}{|c|c|c|c|}
\hline
\multirow{2}{*}{case} & GCC & WeRTC-BBR &Delay-BBR\\ \cline{2-4} 
 & \multicolumn{3}{c|}{(OWD(ms), Loss(\%))} \\ \hline
1 & (118.20, 9.97) &(369.82, 11.59)& (231.51, 1.11) \\ \hline
4 & (117.86, 10.15) &(367.90, 10.06)&(199.11, 0.83)  \\ \hline
7 & (124.93, 11.57) &(355.02, 10.26)&(235.19, 1.23)  \\ \hline
\end{tabular}
\end{table}
\subsection{Performance of the packet schedule algorithm}
\begin{figure}
\centering
\includegraphics[width=3in]{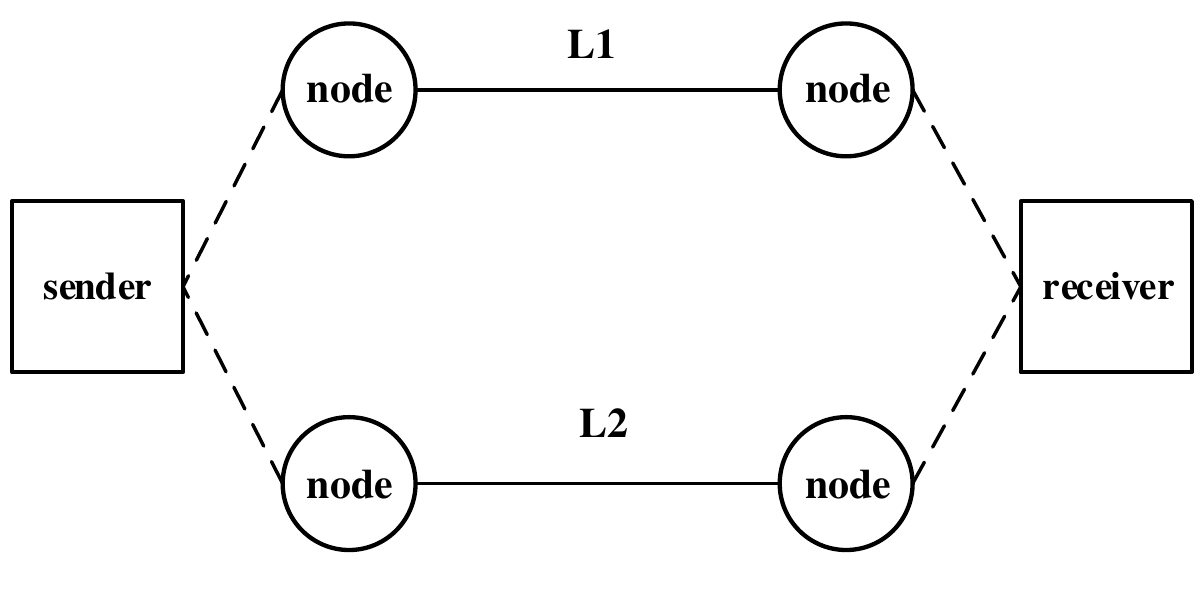}
\caption{Topology}
\label{Fig:topology}
\end{figure}

In this section, the performances of packet scheduling algorithms mentioned above are analyzed. All the packets scheduling algorithms would work under the constraint of the proposed congestion control algorithm. A simple multipath transmission topology was built as shown in Figure \ref{Fig:topology}. Only two paths are tested. The link configuration is shown in Table \ref{Tab:configure}. The parameters on every link are capacity (BW, in unit of Mbps), one way transmission delay (OWD, in unit of milliseconds), link queue length (Q, in unit of milliseconds). When the queued packets exceed the max queue buffer length(Q*BW), the packets loss event will happen due to the working mechanism of Droptail queue management. Totally 10 experiments are designed and all the experiments are running about 300 seconds.
\begin{table}[]
\centering
\caption{Link Configuration}
\label{Tab:configure}
\begin{tabular}{|c|c|c|}
\hline
\multirow{2}{*}{case} & L1                                 & L2                                 \\ \cline{2-3} 
                      & \multicolumn{2}{c|}{(BW, OWD, Q)}                                       \\ \hline
1                     & (4, 100, 200)                      & (4, 100, 200)                            \\ \hline
2                     & (3, 100, 200)                      & (2, 150, 200)                      \\ \hline
3                     & (3, 100, 200)                      & (2, 100, 200)                      \\ \hline
4                     & (4, 100, 200)                      & (2, 50, 200)                      \\ \hline
5                     & (4, 50, 200)                       & (2, 100, 200)                      \\ \hline
6                     & \multicolumn{1}{l|}{(4, 50, 200)} & \multicolumn{1}{l|}{(4, 50, 200)} \\ \hline
7                     & \multicolumn{1}{l|}{(3, 100, 200)} & \multicolumn{1}{l|}{(3, 100, 200)} \\ \hline
8                     & \multicolumn{1}{l|}{(4, 100, 200)} & \multicolumn{1}{l|}{(3, 150, 200)} \\ \hline
9                     & \multicolumn{1}{l|}{(4, 150, 200)} & \multicolumn{1}{l|}{(3, 50, 200)} \\ \hline
10                    & \multicolumn{1}{l|}{(2, 100, 200)} & \multicolumn{1}{l|}{(3, 100, 200)} \\ \hline
\end{tabular}
\end{table}

In consideration that in real network situation, a routing path can be shared by many flows, different flows running on same path to simulated the bandwidth competing situation are tested. Except the multipath session flow, two flows on L1 and one flow on L2 with the same congestion control algorithms are configured in simulation. An ideal video encoder is assumed, which can change the video frame bitrate immediately as requirement of the congestion controller. And a maximum video bitrate 2Mbps is set and the video fame generating bitrate will not be changed if the available bandwidth is above the maximum bitrate. During the simulation process, the frame delay (the interval between the frame received time and generated time) is traced.

\begin{figure}
\centering
\includegraphics[width=3in]{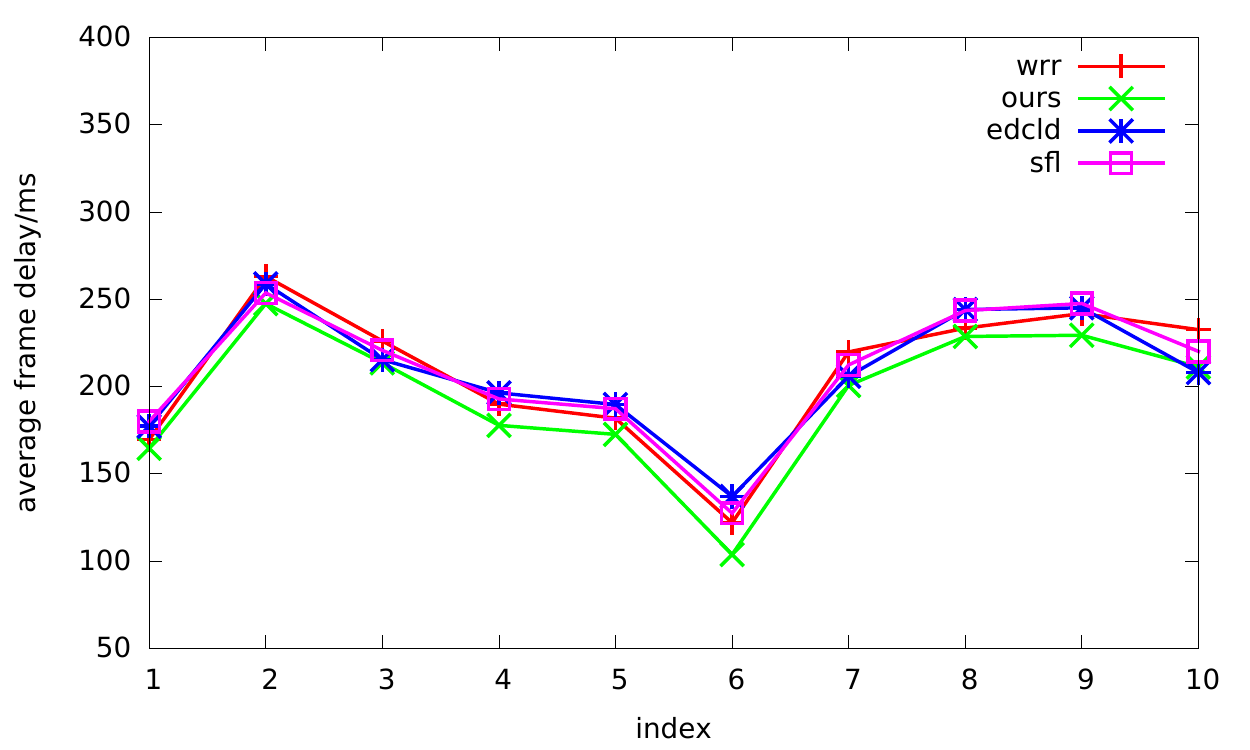}
\caption{The average frame delay}
\label{Fig:frame-delay}
\end{figure}

\begin{figure}
\centering
\includegraphics[width=3in]{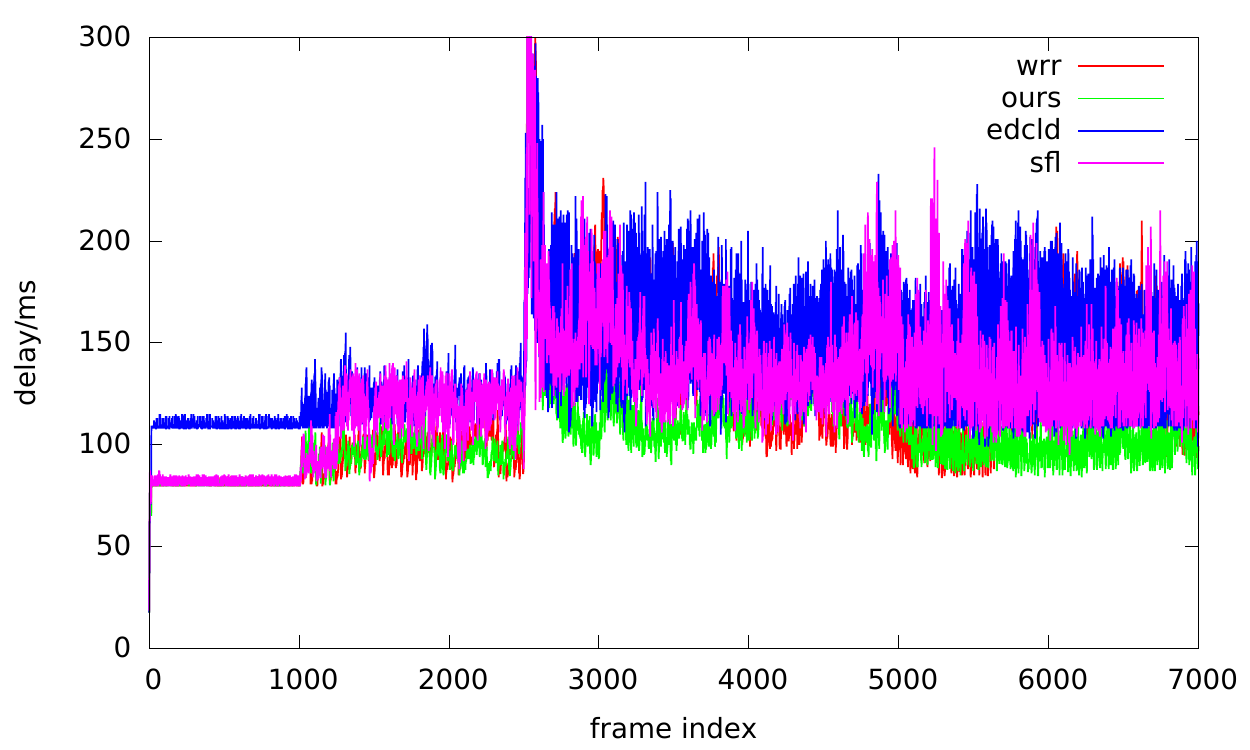}
\caption{The frame delay with link configure 6}
\label{Fig:each-delay}
\end{figure}

There are about 7000 frames during the 300 seconds simulation. The average frame delay of different packets scheduling algorithm are showed in Figure \ref{Fig:frame-delay}. In most test cases, the proposed algorithm achieves lower average frame delay compared the reference schemes. To specifically, each frame arrive delay in test case 6 is shown in Figure \ref{Fig:each-delay}. The delay increase tendency in Figure \ref{Fig:each-delay} is the result of network link congestion when flows competing for bandwidth resource.
\section{Conclusion}
In this paper, a multipath transmission framework for real time video traffic is proposed. Under the multipath transmission context, there are two basic research points: congestion control algorithm and packet scheduling algorithm. The congestion control algorithm is to make the sender maintain a throughput as close as the available bandwidth while avoiding leading the network into congestion status. The most difficult part on designing a congestion control is the fairness property, namely to make the rates of the network user in the same routing path converge to the same throughput with the limited information. And the packet scheduling algorithm is to take advantage of the paths heterogeneity to further improve the performance.

Since the traditional AIMD congestion control is not appropriate for real time video transmission due to the sharp rate reduction in face of link congestion and high transmission delay caused by link buffer occupation, the open source real time video transmission WebRTC implements GCC congestion control algorithm optimized for video transmission on application layer. As simulation results show, there is considerable packet loss rate when GCC flows competing for bandwidth. A modified version of BBR congestion control with delay constraint optimized for video transmission is proposed, which have well fairness property and lower packet. Extensive experiments are conducted to compare the performances of GCC, QUIC-BBR, WebRTC-BBR and our optimized BBR algorithm. Results show that the original BBR implemented in QUIC optimized for bulk data transfer is not quite appropriate for real time traffic, which will cause high packets loss rate when competing bandwidth with other flows with the same congestion control algorithm. To make the BBR congestion control algorithm achieve the claimed optimal control point, in which the sender can achieve the maximum throughput while minimizing end to end latency at the same time, is still an open problem. Even the optimized BBR can achieve lower one way transmission delay than QUIC-BBR and WebRTC-BBR, it is considerable higher when compared with GCC. To further reduce the average transmission delay, which is to reduce the queue occupation in the intermediate routers and maintain the fair bandwidth allocation property is one focus of our future work.

A multipath packet scheduling algorithm induced from utilization maximization theory is proposed. In essence, it can be regarded as earliest first algorithm but takes consideration of local queue buffer length, path available bandwidth and transmission delay. It is compared with WRR, EDCLD, and SFL under the restriction of congestion control algorithm. Simulation results show that the proposed algorithm achieves better performance in term of average frame delay.

Currently, the proposed packet scheduling algorithm only takes queued delay, transmission delay and available bandwidth into consideration. In real network situation, during the process of data transmission, different flows through the same bottleneck link may exploit different congestion control algorithms and some flows over UDP even not implement any congestion control mechanism, thus link congestion is unavoidable and packets loss events happen. For bulk data transfer, the packet loss can be compensated by retransmission. But for real time video traffic, retransmission may not quite effective if the packet arriving time has exceeded the rendering deadline. In future, the combination of multipath scheduling with forward error correction (FEC) to combat the effect of packet loss to further improve the quality of video will be considered.
\section*{References}
\bibliographystyle{elsarticle-num}
\bibliography{mpvideo}

\begin{thebibliography}{10}
\expandafter\ifx\csname url\endcsname\relax
  \def\url#1{\texttt{#1}}\fi
\expandafter\ifx\csname urlprefix\endcsname\relax\def\urlprefix{URL }\fi
\expandafter\ifx\csname href\endcsname\relax
  \def\href#1#2{#2} \def\path#1{#1}\fi

\bibitem{cisco-report}
Cisco,
  \href{http://www.cisco.com/c/en/us/solutions/collateral/service-provider/visual-networking-index-vni/complete-white-paperc11-481360.html}{Cisco
  visual networking index: Forecast and methodology 2013–2018} (Jun. 2014).
\newline\urlprefix\url{http://www.cisco.com/c/en/us/solutions/collateral/service-provider/visual-networking-index-vni/complete-white-paperc11-481360.html}

\bibitem{rfc6356}
C.~Raiciu, M.~Handley, D.~Wischik,
  \href{https://www.rfc-editor.org/info/rfc6356}{Coupled congestion control for
  multipath transport protocols}, RFC 6356, RFC Editor (Oct 2011).
\newline\urlprefix\url{https://www.rfc-editor.org/info/rfc6356}

\bibitem{Khalili2013MPTCP}
R.~Khalili, N.~Gast, M.~Popovic, J.-Y. Le~Boudec,
  \href{http://dx.doi.org/10.1109/TNET.2013.2274462}{Mptcp is not
  pareto-optimal: Performance issues and a possible solution}, IEEE/ACM Trans.
  Netw. 21~(5) (2013) 1651--1665.
\newblock \href {http://dx.doi.org/10.1109/TNET.2013.2274462}
  {\path{doi:10.1109/TNET.2013.2274462}}.
\newline\urlprefix\url{http://dx.doi.org/10.1109/TNET.2013.2274462}

\bibitem{Cao2012Delay}
Y.~Cao, M.~Xu, X.~Fu, Delay-based congestion control for multipath tcp, in:
  2012 20th IEEE International Conference on Network Protocols (ICNP), 2012,
  pp. 1--10.
\newblock \href {http://dx.doi.org/10.1109/ICNP.2012.6459978}
  {\path{doi:10.1109/ICNP.2012.6459978}}.

\bibitem{Frommgen2016ReMP}
A.~Frommgen, T.~Erbshäußer, A.~Buchmann, T.~Zimmermann, K.~Wehrle, Remp tcp:
  Low latency multipath tcp, in: 2016 IEEE International Conference on
  Communications (ICC), 2016, pp. 1--7.
\newblock \href {http://dx.doi.org/10.1109/ICC.2016.7510787}
  {\path{doi:10.1109/ICC.2016.7510787}}.

\bibitem{Guo2017Accelerating}
Y.~E. Guo, A.~Nikravesh, Z.~M. Mao, F.~Qian, S.~Sen,
  \href{http://doi.acm.org/10.1145/3117811.3117829}{Accelerating multipath
  transport through balanced subflow completion}, in: Proceedings of the 23rd
  Annual International Conference on Mobile Computing and Networking, MobiCom
  '17, ACM, New York, NY, USA, 2017, pp. 141--153.
\newblock \href {http://dx.doi.org/10.1145/3117811.3117829}
  {\path{doi:10.1145/3117811.3117829}}.
\newline\urlprefix\url{http://doi.acm.org/10.1145/3117811.3117829}

\bibitem{Lim2017ECF}
Y.-s. Lim, E.~M. Nahum, D.~Towsley, R.~J. Gibbens,
  \href{http://doi.acm.org/10.1145/3143314.3078552}{Ecf: An mptcp path
  scheduler to manage heterogeneous paths}, SIGMETRICS Perform. Eval. Rev.
  45~(1) (2017) 33--34.
\newblock \href {http://dx.doi.org/10.1145/3143314.3078552}
  {\path{doi:10.1145/3143314.3078552}}.
\newline\urlprefix\url{http://doi.acm.org/10.1145/3143314.3078552}

\bibitem{Shi2018STMS}
H.~Shi, Y.~Cui, X.~Wang, Y.~Hu, M.~Dai, F.~Wang, K.~Zheng, Stms: Improving
  mptcp throughput under heterogeneous networks, in: 2018 USENIX Annual
  Technical Conference, 2018, pp. 719--730.

\bibitem{Shreedhar2018QAware}
T.~Shreedhar, N.~Mohan, S.~K. Kaul, J.~Kangasharju, Qaware: A cross-layer
  approach to mptcp scheduling, arXiv preprint arXiv:1808.04390.

\bibitem{Ferlin2016BLEST}
S.~Ferlin, O.~Alay, O.~Mehani, R.~Boreli, Blest: Blocking estimation-based
  mptcp scheduler for heterogeneous networks, in: 2016 IFIP Networking
  Conference (IFIP Networking) and Workshops, 2016, pp. 431--439.
\newblock \href {http://dx.doi.org/10.1109/IFIPNetworking.2016.7497206}
  {\path{doi:10.1109/IFIPNetworking.2016.7497206}}.

\bibitem{jacobson1988congestion}
V.~Jacobson, Congestion avoidance and control, in: ACM SIGCOMM computer
  communication review, Vol.~18, ACM, 1988, pp. 314--329.

\bibitem{cardwell2016bbr}
N.~Cardwell, Y.~Cheng, C.~S. Gunn, S.~H. Yeganeh, V.~Jacobson, Bbr:
  Congestion-based congestion control, Queue 14~(5) (2016) 50.

\bibitem{ietf-rmcat-require}
E.~Rescorla,
  \href{https://tools.ietf.org/html/draft-ietf-rmcat-cc-requirements-09}{Congestion
  control requirements for rmcat}, Internet-draft, Internet Engineering Task
  Force, work in Progress (Dec. 2014).
\newline\urlprefix\url{https://tools.ietf.org/html/draft-ietf-rmcat-cc-requirements-09}

\bibitem{xu2004binary}
L.~Xu, K.~Harfoush, I.~Rhee, Binary increase congestion control (bic) for fast
  long-distance networks, in: INFOCOM 2004. Twenty-third AnnualJoint Conference
  of the IEEE Computer and Communications Societies, Vol.~4, IEEE, 2004, pp.
  2514--2524.

\bibitem{ha2008cubic}
S.~Ha, I.~Rhee, L.~Xu, Cubic: a new tcp-friendly high-speed tcp variant, ACM
  SIGOPS operating systems review 42~(5) (2008) 64--74.

\bibitem{Staff2012Bufferbloat}
C.~Staff, Bufferbloat: What's wrong with the internet?, Communications of the
  ACM 55~(2) (2012) 40--47.

\bibitem{carlucci2016analysis}
G.~Carlucci, L.~De~Cicco, S.~Holmer, S.~Mascolo, Analysis and design of the
  google congestion control for web real-time communication (webrtc), in:
  Proceedings of the 7th International Conference on Multimedia Systems, ACM,
  2016, p.~13.

\bibitem{zhu2013nada}
X.~Zhu, R.~Pan, M.~Ramalho, S.~de~la Cruz, C.~Ganzhorn, P.~Jones,
  S.~D’Aronco,
  \href{https://tools.ietf.org/html/draft-ietf-rmcat-nada-07}{Nada: A unified
  congestion control scheme for real-time media}, Internet-draft, Internet
  Engineering Task Force, work in Progress (2018).
\newline\urlprefix\url{https://tools.ietf.org/html/draft-ietf-rmcat-nada-07}

\bibitem{rfc8298}
I.~Johansson, Z.~Sarker,
  \href{https://www.rfc-editor.org/rfc/rfc8298.txt}{Self-clocked rate
  adaptation for multimedia}, RFC 8298, RFC Editor (Dec 2017).
\newline\urlprefix\url{https://www.rfc-editor.org/rfc/rfc8298.txt}

\bibitem{Kuhn2014DAPS}
N.~Kuhn, E.~Lochin, A.~Mifdaoui, G.~Sarwar, O.~Mehani, R.~Boreli, Daps:
  Intelligent delay-aware packet scheduling for multipath transport, in: 2014
  IEEE International Conference on Communications (ICC), 2014, pp. 1222--1227.
\newblock \href {http://dx.doi.org/10.1109/ICC.2014.6883488}
  {\path{doi:10.1109/ICC.2014.6883488}}.

\bibitem{Cetinkaya2004Opportunistic}
C.~Cetinkaya, E.~W. Knightly, Opportunistic traffic scheduling over multiple
  network paths, in: IEEE INFOCOM 2004, Vol.~3, 2004, pp. 1928--1937 vol.3.
\newblock \href {http://dx.doi.org/10.1109/INFCOM.2004.1354602}
  {\path{doi:10.1109/INFCOM.2004.1354602}}.

\bibitem{Ribeiro2003pathchirp}
V.~J. Ribeiro, R.~H. Riedi, R.~G. Baraniuk, J.~Navratil, L.~Cottrell,
  pathchirp: Efficient available bandwidth estimation for network paths, in:
  Passive and active measurement workshop, 2003.

\bibitem{Norros1994storage}
I.~Norros, \href{https://doi.org/10.1007/BF01158964}{A storage model with
  self-similar input}, Queueing Systems 16~(3) (1994) 387--396.
\newblock \href {http://dx.doi.org/10.1007/BF01158964}
  {\path{doi:10.1007/BF01158964}}.
\newline\urlprefix\url{https://doi.org/10.1007/BF01158964}

\bibitem{Zhu2007Rate}
X.~Zhu, P.~Agrawal, J.~Pal~Singh, T.~Alpcan, B.~Girod,
  \href{http://doi.acm.org/10.1145/1291233.1291247}{Rate allocation for
  multi-user video streaming over heterogenous access networks}, in:
  Proceedings of the 15th ACM International Conference on Multimedia, MM '07,
  ACM, New York, NY, USA, 2007, pp. 37--46.
\newblock \href {http://dx.doi.org/10.1145/1291233.1291247}
  {\path{doi:10.1145/1291233.1291247}}.
\newline\urlprefix\url{http://doi.acm.org/10.1145/1291233.1291247}

\bibitem{Bui2010Markovian}
V.~Bui, W.~Zhu, A.~Botta, A.~Pescape, A markovian approach to multipath data
  transfer in overlay networks, IEEE Transactions on Parallel and Distributed
  Systems 21~(10) (2010) 1398--1411.
\newblock \href {http://dx.doi.org/10.1109/TPDS.2010.13}
  {\path{doi:10.1109/TPDS.2010.13}}.

\bibitem{Prabhavat2011Effective}
S.~Prabhavat, H.~Nishiyama, N.~Ansari, N.~Kato, Effective delay-controlled load
  distribution over multipath networks, IEEE Transactions on Parallel and
  Distributed Systems 22~(10) (2011) 1730--1741.
\newblock \href {http://dx.doi.org/10.1109/TPDS.2011.43}
  {\path{doi:10.1109/TPDS.2011.43}}.

\bibitem{Wu2015low}
J.~Wu, X.~Qiao, Y.~Xia, C.~Yuen, J.~Chen,
  \href{https://doi.org/10.1007/s00530-014-0388-7}{A low-latency scheduling
  approach for high-definition video streaming in a heterogeneous wireless
  network with multihomed clients}, Multimedia Systems 21~(4) (2015) 411--425.
\newblock \href {http://dx.doi.org/10.1007/s00530-014-0388-7}
  {\path{doi:10.1007/s00530-014-0388-7}}.
\newline\urlprefix\url{https://doi.org/10.1007/s00530-014-0388-7}

\bibitem{Aggarwal2000Understanding}
A.~Aggarwal, S.~Savage, T.~Anderson, Understanding the performance of tcp
  pacing, in: Proceedings IEEE INFOCOM 2000. Conference on Computer
  Communications. Nineteenth Annual Joint Conference of the IEEE Computer and
  Communications Societies, IEEE, 2000, pp. 1157--1165.

\bibitem{Wei2006TCP}
D.~Wei, P.~Cao, S.~Low, C.~EAS, Tcp pacing revisited 2 (2006) 3.

\bibitem{holmer2015rtp}
F.~S, Holmerand~M, S.~E,
  \href{https://tools.ietf.org/html/draft-holmer-rmcat-transport-wide-cc-extensions-01}{Rtp
  extensions for transport-wide congestion control}, Internet-draft, Internet
  Engineering Task Force, work in Progress (Oct. 2015).
\newline\urlprefix\url{https://tools.ietf.org/html/draft-holmer-rmcat-transport-wide-cc-extensions-01}

\bibitem{Kelly1998Rate}
F.~P. Kelly, A.~K. Maulloo, D.~K.~H. Tan,
  \href{https://doi.org/10.1057/palgrave.jors.2600523}{Rate control for
  communication networks: shadow prices, proportional fairness and stability},
  Journal of the Operational Research Society 49~(3) (1998) 237--252.
\newblock \href {http://dx.doi.org/10.1057/palgrave.jors.2600523}
  {\path{doi:10.1057/palgrave.jors.2600523}}.
\newline\urlprefix\url{https://doi.org/10.1057/palgrave.jors.2600523}

\bibitem{Low1999Optimization}
S.~H. Low, D.~E. Lapsley, Optimization flow control. i. basic algorithm and
  convergence, IEEE/ACM Transactions on Networking 7~(6) (1999) 861--874.
\newblock \href {http://dx.doi.org/10.1109/90.811451}
  {\path{doi:10.1109/90.811451}}.

\bibitem{Paganini2009Unified}
F.~Paganini, E.~Mallada, A unified approach to congestion control and
  node-based multipath routing, IEEE/ACM Transactions on Networking 17~(5)
  (2009) 1413--1426.
\newblock \href {http://dx.doi.org/10.1109/TNET.2008.2011902}
  {\path{doi:10.1109/TNET.2008.2011902}}.

\bibitem{Zhang2018Congestion}
S.~Zhang, Congestion control for rtp media: a comparison on simulated
  environment, arXiv preprint arXiv:1809.00304.

\end{thebibliography}
\end{document}